\documentclass[10pt,a4paper]{article}

\usepackage[T1]{fontenc}
\usepackage[utf8]{inputenc}
\usepackage{amsmath,amssymb,amsfonts,amsthm}
\usepackage{newtxtext,newtxmath}
\usepackage{microtype}
\usepackage[a4paper,top=1.55cm,bottom=1.70cm,left=1.62cm,right=1.62cm,columnsep=0.64cm]{geometry}
\usepackage{dblfloatfix}

\usepackage{bm}
\usepackage{algorithm}
\usepackage{algorithmicx}
\usepackage{algpseudocode}

\usepackage{graphicx}
\graphicspath{{figures/}{./}}
\usepackage{xcolor}
\usepackage{booktabs}
\usepackage{multirow}
\usepackage{array}
\usepackage{tabularx}
\usepackage{adjustbox}
\usepackage{overpic}
\usepackage{caption}
\usepackage{subcaption}
\usepackage{balance}
\usepackage{placeins}

\usepackage[inline]{enumitem}
\usepackage{titlesec}
\usepackage{fancyhdr}
\usepackage{comment}
\usepackage{textcomp}
\usepackage{listings}
\usepackage[numbers,sort&compress]{natbib}
\usepackage{xurl}
\usepackage{hyperref}
\usepackage{orcidlink}

\hypersetup{
  pdftitle={MDQEC-QAS: Meta-Decoding for Quantum Error Correction with Hardware-Aware VQC Search and Confidence-Gated Recovery},
  pdfauthor={Prashant Kumar Choudhary, Nouhaila Innan, Muhammad Shafique, Rajeev Singh},
  pdfsubject={Quantum error correction, meta-decoding, variational quantum circuits, and quantum architecture search},
  pdfkeywords={quantum error correction, meta-decoding, variational quantum circuits, quantum architecture search, stabilizer codes, planar codes, transfer learning, hybrid decoding},
  colorlinks=true,
  linkcolor=black,
  citecolor=blue!55!black,
  urlcolor=blue!55!black
}

\setlist[itemize]{leftmargin=*,topsep=3pt,itemsep=2pt,parsep=0pt}
\setlist[enumerate]{leftmargin=*,topsep=3pt,itemsep=2pt,parsep=0pt}

\titleformat{\section}
  {\large\bfseries\raggedright}
  {\thesection}{0.55em}{}
\titlespacing*{\section}{0pt}{9pt plus 2pt minus 1pt}{4pt}
\titleformat{\subsection}
  {\normalsize\bfseries\raggedright}
  {\thesubsection}{0.5em}{}
\titlespacing*{\subsection}{0pt}{7pt plus 2pt minus 1pt}{3pt}

\fancypagestyle{plain}{%
  \fancyhf{}
  \fancyhead[L]{\footnotesize MDQEC-QAS}
  \fancyhead[R]{\footnotesize Choudhary et al.}
  \fancyfoot[C]{\footnotesize\thepage}
  
}

\AtBeginDocument{}

\begin{document}

\twocolumn[
\begin{@twocolumnfalse}
\begin{center}
{\LARGE\bfseries MDQEC-QAS: Meta-Decoding for Quantum Error Correction with Hardware-Aware VQC Search and Confidence-Gated Recovery\par}
\vspace{0.70em}
{\large
Prashant Kumar Choudhary\textsuperscript{1}\,\orcidlink{0009-0005-9912-3231},
Nouhaila Innan\textsuperscript{2,3}\,\orcidlink{0000-0002-1014-3457},
Muhammad Shafique\textsuperscript{2,3}\,\orcidlink{0000-0002-2607-8135} and
Rajeev Singh\textsuperscript{1}\,\orcidlink{0000-0002-0768-6549}\par}
\vspace{0.45em}
{\small
\textsuperscript{1}Department of Physics, Indian Institute of Technology (BHU), Varanasi, Uttar Pradesh, India\\
\textsuperscript{2}eBRAIN Lab, Division of Engineering, New York University Abu Dhabi (NYUAD), Abu Dhabi, UAE\\
\textsuperscript{3}Center for Quantum and Topological Systems (CQTS), NYUAD Research Institute, NYUAD, Abu Dhabi, UAE\par}
\vspace{0.35em}
{\footnotesize
\href{mailto:prashantkchoudhary.rs.phy22@iitbhu.ac.in}{prashantkchoudhary.rs.phy22@iitbhu.ac.in}\quad
\href{mailto:nouhaila.innan@nyu.edu}{nouhaila.innan@nyu.edu}\quad
\href{mailto:muhammad.shafique@nyu.edu}{muhammad.shafique@nyu.edu}\quad
\href{mailto:rajeevs.phy@iitbhu.ac.in}{rajeevs.phy@iitbhu.ac.in}}
\end{center}

\vspace{0.65em}
\noindent\begin{minipage}{0.96\textwidth}
\small
\textbf{Abstract.}
We propose a unified meta-decoding framework for quantum error correction that learns syndrome-to-recovery mappings across multiple stabilizer codes and noise settings, without separate decoders for each configuration. The benchmark includes FiveQubit, Steane, Planar3$\times$3 and Planar5$\times$5 codes, four noise families, and five regimes: interpolation, unseen-\(p\) transfer, unseen-noise transfer, few-shot unseen-code adaptation, and few-shot held-out-size adaptation. We compare a classical Meta-MLP teacher-trained baseline with variational quantum circuit (VQC) meta-decoders selected through hardware-aware quantum architecture search over qubit count, circuit depth, and entangling topology. The Meta-MLP achieves teacher-label accuracies of 0.9993, 0.9118, 0.9342, 0.6304, and 0.7548 across the five regimes, while the hardware-aware VQC achieves 0.9400, 0.8495, 0.8415, 0.5678, and 0.7143. However, logical-level evaluation shows that high teacher-label accuracy alone is insufficient in the hardest Planar5$\times$5 setting. During interpolation, raw logical-failure ratios relative to the teacher are 12.08 and 25.91 for the Meta-MLP and VQC, respectively, but confidence-gated fallback reduces them to 1.71 and 1.11. These results support confidence-aware selective recovery rather than unconditional teacher replacement.

\vspace{0.45em}
\end{minipage}
\vspace{0.75em}
\hrule
\vspace{1.0em}
\end{@twocolumnfalse}
]

\section{Introduction}
\label{sec:intro}

Quantum error correction (QEC) is a central requirement for scalable fault-tolerant quantum computation because fragile quantum states are continuously exposed to decoherence, control imperfections, and measurement noise. Within this landscape, the stabilizer formalism provides a unifying framework for constructing and analyzing a broad class of quantum codes, including both compact algebraic codes and large topological codes \cite{Shor1995Scheme,Steane1996MultipleParticle,Gottesman1997Stabilizer,Kitaev2003FaultTolerant,Terhal2015QECReview}. In a stabilizer-code setting, noisy physical qubits are probed through stabilizer measurements, producing a classical syndrome that must be translated into a recovery operation. The success of this decoding stage directly determines the resulting logical failure probability and therefore the usefulness of the encoded quantum information \cite{Dennis2002TopologicalMemory,Fowler2012SurfaceCode,Campbell2017RoadsToFaultTolerance}.

For a measured syndrome \(s\), decoding can be considered abstractly as the inference problem
\begin{equation}
r^{\star}(s)=\arg\max_{r}\Pr(r\mid s),
\label{eq:intro_map}
\end{equation}
where \(r\) denotes a candidate recovery operator. The operational metric of interest is the logical failure probability
\begin{equation}
p_{L}(p)=\Pr(\text{logical failure}\mid p),
\label{eq:intro_pl}
\end{equation}
where \(p\) is the physical error probability or, more generally, a parameter controlling the noise strength. In other words, a useful decoder is not just one that predicts labels accurately, but one that preserves logical information after recovery. This distinction becomes particularly important for topological codes, where two decoders with similar classification accuracy can still differ substantially in logical performance \cite{deMartiOlius2024SurfaceCodeReview}.

Traditional decoders such as lookup-table strategies, heuristic code-specific rules, and matching-based methods can perform strongly within narrowly defined settings, but they are usually specialized to a certain code family or noise model \cite{Delfosse2020DecodersReview,Chamberland2023MLDecodingSurvey}. This limitation has motivated machine-learning-based decoders, from early neural approaches to more scalable graph- and transformer-style models \cite{Bausch2024LearningDecoder,z1ng-wg3k}.

These developments establish two key observations. First, learned decoders can be competitive with hand-engineered approaches, particularly when the noise has correlations or when decoder inference must adapt to experimental data \cite{Baireuther2018CorrelatedDecoder}. Second, the most challenging regimes are often larger structured topological codes, where raw classification accuracy alone does not guarantee strong logical performance 
\cite{deMartiOlius2024SurfaceCodeReview,Fischer2023HardnessRF}. Thus, modern QEC decoding is no longer only a combinatorial problem; it increasingly also involves representation learning and decoder-system design.

Despite this progress, most learned QEC decoders are still trained and evaluated within fixed code/noise regimes. However, a robust decoder should ideally adapt across heterogeneous physical conditions: different stabilizer structures, different noise families, and different physical error rates. This motivates a \emph{meta-decoding} perspective, in which one shared model is trained across multiple decoding tasks and receives explicit side information describing the code family, the noise type, and the physical error level. Rather than learning one isolated decoder per setting, the model learns a transferable syndrome-to-recovery representation \cite{Hammar2022ErrorRateAgnostic,PhysRevResearch.7.013029}.

In parallel, variational quantum circuits (VQCs) have emerged as a candidate for hybrid quantum-classical learning systems, but their performance depends strongly on ansatz design \cite{Havlicek2019QuantumFeatureSpaces,Schuld2020CircuitCentric}. This makes quantum architecture search (QAS) a relevant tool for QEC decoding as well: instead of fixing a single parametrized quantum circuit by hand, one may search over qubit count, circuit depth, and entangling topology, and then select architectures using both predictive quality and implementation-oriented cost proxies \cite{Du2022QCAS,Martyniuk2024QASSurvey,choudhary2025graphbasedbayesianoptimizationquantum,marchisio2026hybrid,kashif2026faqnas,dutta2025qas}. In a decoding pipeline, this leads naturally to a hardware-aware model-selection problem rather than a purely accuracy-driven one.

In this work, we propose a unified meta-decoding framework for QEC that combines these ideas in a single reproducible pipeline. The framework jointly studies two algebraic codes (FiveQubit and Steane), two planar topological codes (Planar3$\times$3 and Planar5$\times$5), four noise families (depolarizing, \(Z\)-biased, correlated \(Z\)-burst, and depolarizing with syndrome-measurement flips), and several transfer settings, including interpolation, unseen-\(p\) transfer, unseen-noise transfer, and few-shot adaptation to an unseen code. We compare a classical meta-MLP against VQC meta-decoders selected through hardware-aware QAS.

A key observation arising from our experiments is that the larger planar code, Planar5$\times$5, is substantially more sensitive to raw learned-decoder mistakes at the logical level than the smaller codes. Even rare teacher-label errors can become logically costly, so high supervised accuracy alone is not a sufficient reliability criterion for this regime. To address this, we introduce a confidence-gated hybrid recovery strategy: when the learned decoder is sufficiently confident, its prediction is used directly; otherwise, the pipeline falls back to the teacher decoder. This mechanism reveals a high-precision selective-decoding regime: the learned model can be useful on confident cases even when it is not reliable as a full teacher replacement. The resulting system should
therefore be interpreted as a hybrid learned-assisted recovery rather than an independent learned decoding. Confidence-aware and risk-aware mechanisms are standard in broader machine learning and decision systems, but they remain underexplored in QEC decoding pipelines \cite{Pan2024AIQEC}.

The novelty of this work does not lie in the isolated use of a neural decoder or a variational quantum classifier. Rather, it lies in the integration of four components within a single reproducible QEC pipeline: (i) multi-code, multi-noise meta-decoding across algebraic and planar stabilizer codes, (ii) transfer evaluation under unseen-$p$, unseen-noise, few-shot unseen-code, and held-out-size settings, (iii) compact hardware-aware quantum architecture search for VQC decoder design, and (iv) confidence-gated hybrid recovery for the most challenging topological-code regime. This integrated design leads to three system-level findings: confidence can act as a routing signal for selective decoding, structural transfer across code families and code sizes is substantially harder than transfer across noise parameters, and cost-aware VQC selection can remain competitive with accuracy-selected circuits within the compact search space.

The main contributions of this paper are summarized as follows:
\begin{itemize}
\item We construct a pooled multi-code, multi-noise QEC meta-decoding benchmark spanning algebraic and planar stabilizer codes, multiple noise families, and several transfer settings.
\item We compare a classical Meta-MLP with hardware-aware VQC meta-decoders under a shared global recovery-label framework.
\item We perform compact hardware-aware QAS over qubit count, circuit depth, and entangling topology, and evaluate the selected VQC architectures after full retraining.
\item We distinguish teacher-label accuracy from logical-level decoding performance, showing that raw supervised accuracy can be misleading for larger topological codes.
\item We identify a transfer hierarchy in which unseen-$p$ and unseen-noise generalization are easier than unseen-code and held-out-size adaptation, indicating that changes in code structure are more challenging than changes in the noise distribution.
\item We introduce and evaluate confidence-gated hybrid recovery, showing that confidence can serve as a routing signal between learned predictions and teacher fallback in the larger Planar5$\times$5 regime.
\end{itemize}

\section{Background and Related Work}
\label{sec:related-work}

Research on quantum error-correction decoders has evolved from analytical and combinatorial constructions toward increasingly data-driven and learning-based approaches. This section reviews prior work at the intersection of stabilizer-code decoding, learned decoder transferability, variational quantum models, and hardware-aware decoder design.

\subsection{Classical and learned decoding for stabilizer codes}

Quantum decoding has traditionally been studied through code-specific analytical and combinatorial methods. Within the stabilizer formalism, the decoder receives a syndrome and must infer a recovery that returns the state to the codespace without inducing a logical fault \cite{Gottesman1997Stabilizer,Terhal2015QECReview}. For topological and surface-code families, decoding is closely connected to threshold analysis and fault-tolerant architecture design. In particular, the topological-quantum-memory framework and later surface-code studies established minimum-weight matching and related combinatorial approaches as strong baselines for structured local-noise settings \cite{Dennis2002TopologicalMemory,Fowler2012SurfaceCode,deMartiOlius2024SurfaceCodeReview}. These decoders are often highly effective, but they are typically specialized to a particular code structure and noise assumption \cite{Berent_2024}.

This specialization motivates the search for more adaptive decoder families, since modern experimental platforms exhibit changing calibration, asymmetric noise, measurement faults, and correlated errors. As a result, decoder design increasingly requires methods that combine combinatorial optimization, representation learning, and system-level design.


This need for adaptability helped motivate machine-learning-based decoders as data-driven alternatives to explicit code-specific decoding rules. Torlai and Melko provided one of the earliest demonstrations that neural networks can decode topological codes by learning syndrome-to-recovery mappings directly from data \cite{Torlai2017NeuralDecoder}. Shortly afterward, Varsamopoulos \emph{et al.} showed that feedforward neural networks can decode small surface codes efficiently while accounting for implementation-oriented decoding latency \cite{Varsamopoulos2018FeedforwardDecoder}. These studies established that decoder behavior can be approximated by trainable models rather than relying exclusively on explicit combinatorial procedures.

Subsequent work extended learned decoding to more structured regimes. Baireuther \emph{et al.} studied machine-learning-assisted correction of correlated qubit errors in topological codes, showing that learned approaches can exploit correlations that may be difficult to incorporate into conventional decoders \cite{Baireuther2018CorrelatedDecoder}. Reinforcement-learning-based approaches further explored adaptive decoder behavior and control strategies \cite{Andreasson2019RLDecoder}. More recently, error-rate-agnostic decoding studies investigated whether neural decoders can remain robust when the physical error rate differs between training and testing, thereby motivating transfer-oriented evaluations in QEC \cite{Hammar2022ErrorRateAgnostic,PhysRevResearch.7.013029}.

These prior works established the feasibility of learned decoding, but most remained centered on a single code family, a single lattice class, or a single noise scenario. In contrast, this study examines a unified meta-decoding framework spanning multiple stabilizer codes, multiple noise families, and explicit transfer settings.

\subsection{Structured decoders, transferability, and logical-level reliability}

Recent learned QEC decoders have increasingly adopted structured architectures, including graph neural networks and transformer-style models. This development is motivated by the locality and connectivity patterns of many quantum codes, especially topological codes, which are not always captured efficiently by flat multilayer perceptrons. Lange \emph{et al.} demonstrated graph-neural-network-based decoding of quantum error-correcting codes, showing that structure-aware learned models can provide strong decoding performance in more complex regimes \cite{Lange2025DataDriven}. Similarly, recent large-scale work on processor data showed that high-capacity learned decoders can achieve strong decoding quality under experimental noise, highlighting the importance of expressive architectures and training distributions \cite{Bausch2024LearningDecoder,z1ng-wg3k}.

These developments are closely related to the motivation of this study, but the present focus differs in two ways. First, rather than optimizing a learned decoder for a single code family, this study considers a \emph{multi-code, multi-noise} setting in which a shared decoder is conditioned on the code family, noise family, and physical error rate. Second, this study examines the gap between \emph{classification performance} and \emph{logical-level performance}, especially for the larger planar code. This distinction is important because raw learned decoders may achieve nontrivial classification accuracy while still producing substantial logical degradation unless a hybrid recovery mechanism is used \cite{Fischer2023HardnessRF}.


Beyond architectural design, transferability has become a central concern for learned decoding. A major open question in QEC machine learning is whether learned decoders can transfer beyond the regime on which they were trained. Recent error-rate-agnostic and near-term surface-code decoding studies have highlighted this issue by showing that decoders trained under one physical-error regime may generalize imperfectly when evaluated under another \cite{Hammar2022ErrorRateAgnostic,PhysRevResearch.7.013029}.

This study extends the transfer perspective by evaluating generalization across interpolation, unseen-$p$, unseen-noise, few-shot unseen-code, and held-out-size settings. This combination remains relatively uncommon in prior decoder work, particularly when classical and VQC decoders are evaluated within a unified framework.

A recurring challenge in learned QEC decoding is that strong label-level prediction does not necessarily imply strong logical-level recovery, particularly for larger structured codes. This distinction is central to the present study: raw learned decoders, including both the classical meta-MLP and the VQC models, can degrade on Planar5$\times$5 even when their supervised accuracy remains nontrivial. This motivates a hybrid recovery formulation in which learned inference is used selectively and low-confidence cases are delegated to a teacher decoder \cite{Pan2024AIQEC}.

Although confidence-based hybridization is common in broader machine-learning systems, it has been less explicitly studied as part of QEC decoder pipelines. In this study, confidence-gated hybrid recovery is used to connect learned transferability with logical reliability, especially in the larger planar-code regime.
\subsection{Variational quantum models and architecture search}

Variational quantum circuits have become an important model class in quantum machine learning. Quantum-enhanced feature spaces \cite{Havlicek2019QuantumFeatureSpaces}, circuit-centric quantum classifiers \cite{Schuld2020CircuitCentric}, data re-uploading architectures \cite{PerezSalinas2020DataReuploading}, and quantum convolutional neural networks \cite{Cong2019QCNN} demonstrate that trainable quantum circuits can serve as expressive learning models. These models have been studied across a range of applications, including classification, regression, generative modeling, optimization, and domain-specific prediction tasks, often in hybrid quantum-classical pipelines designed for near-term quantum devices \cite{innan2024quantum,innan2025qnn,innan2025lep,Choudhary2026HQNNFSP}. At the same time, their performance depends strongly on circuit architecture, including qubit count, circuit depth, entangling pattern, parameter initialization, and trainability \cite{Skolik2021,Cerezo2021,vyskubov2026scaling}.

This dependence has motivated quantum architecture search. Du \emph{et al.} studied quantum circuit architecture search for variational quantum algorithms \cite{Du2022QCAS}, while later work and surveys examined broader directions in automated quantum model design \cite{Martyniuk2024QASSurvey,7rc4-p446}. However, most of this literature does not focus specifically on QEC decoding, and relatively little work addresses transfer-oriented multi-code decoder design. This gap motivates hardware-aware QAS for QEC meta-decoding, with circuit architectures evaluated using predictive performance and implementation-oriented cost proxies such as depth and two-qubit gate count.

\begin{table*}[t]
\centering
\caption{Positioning of this study relative to prior directions in learned QEC decoding.}
\label{tab:related_positioning}
\small
\begin{tabular}{p{0.22\textwidth}p{0.23\textwidth}p{0.49\textwidth}}
\toprule
\textbf{Prior direction} & \textbf{Scope} & \textbf{Distinction of this study} \\
\midrule
Early neural decoding \cite{Torlai2017NeuralDecoder,Varsamopoulos2018FeedforwardDecoder} &
Single code and fixed noise model &
Multi-code, multi-noise meta-decoding across algebraic and planar stabilizer codes \\

Correlated and adaptive learned decoding \cite{Baireuther2018CorrelatedDecoder,Andreasson2019RLDecoder,Hammar2022ErrorRateAgnostic} &
Topological codes with correlated noise or varying error rates &
Explicit evaluation under unseen-$p$, unseen-noise, and few-shot adaptation settings \\

Graph-based and large-scale learned decoding \cite{Lange2025DataDriven,Bausch2024LearningDecoder,z1ng-wg3k} &
Larger or more structured topological-code settings &
Unified evaluation across multiple code families, including both prediction accuracy and logical-level performance \\

Quantum-model-based learning \cite{Havlicek2019QuantumFeatureSpaces,Schuld2020CircuitCentric,Cong2019QCNN} &
VQC, QCNN, and circuit-based learning models &
Hardware-aware VQC architecture search applied specifically to QEC meta-decoding \\
\midrule
\textbf{Our proposed framework} &
Multi-code, multi-noise, transfer-oriented, and logical-level QEC decoding &
Confidence-gated hybrid recovery for larger topological-code regimes, with teacher fallback in low-confidence cases \\
\bottomrule
\end{tabular}
\end{table*}
\subsection{Positioning within prior QEC decoding studies} \label{subsec:positioning-present-work}

Our study builds on these prior directions while differing in scope, evaluation design, and decoder-system formulation. Compared with early neural decoders, the proposed framework moves beyond single-code decoding toward a shared meta-decoder trained across multiple code and noise settings. Compared with correlated-noise and error-rate-adaptive learned decoders, it evaluates transfer across both noise distributions and code structures. Compared with recent graph-based and large-scale learned decoders, its focus is not limited to decoder accuracy, but also includes the relationship between supervised prediction quality and logical-level recovery. Finally, relative to general QAS studies, this work applies hardware-aware VQC search specifically to QEC meta-decoding.

Accordingly, our work is positioned as an integrated QEC decoding framework that combines multi-code meta-decoding, transfer evaluation, hardware-aware VQC model selection, and confidence-gated hybrid recovery. Table~\ref{tab:related_positioning} makes this distinction explicit by contrasting the present framework with representative prior directions in learned QEC decoding.

\section{Methodology}
\label{sec:methodology}

\begin{figure*}[t]
\centering
\includegraphics[width=0.96\textwidth]{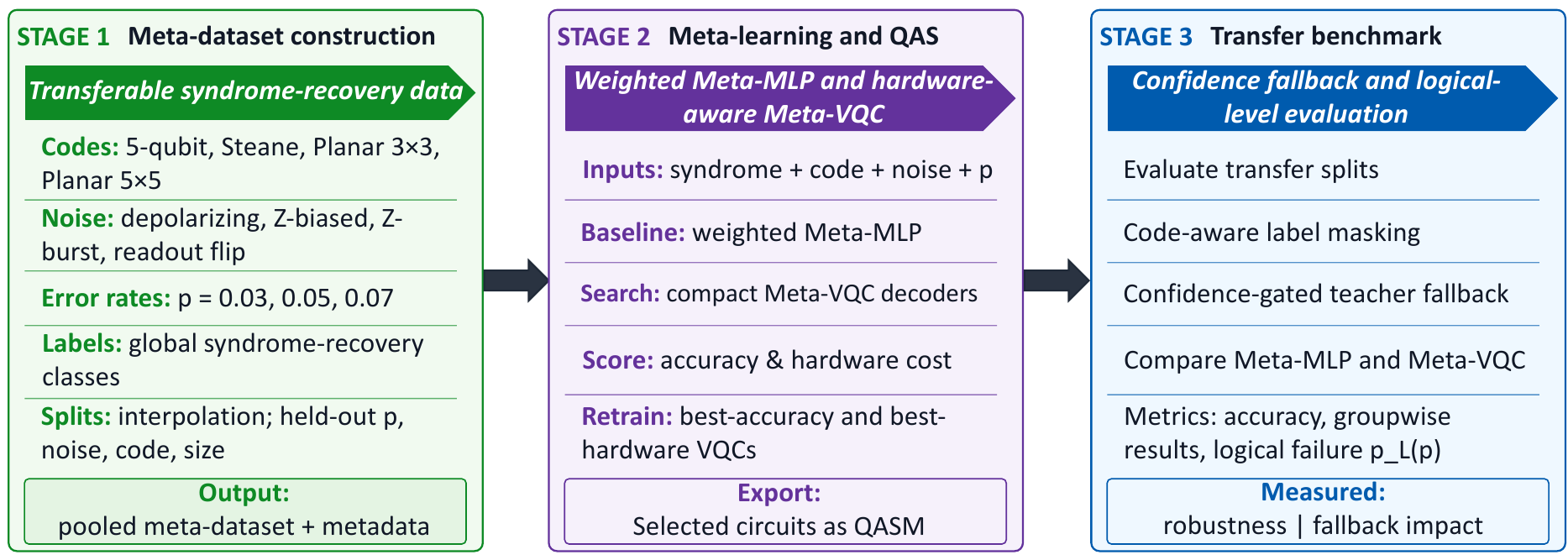}
\caption{Overview of the learned-assisted meta-decoding framework. Syndrome-recovery data are generated using code-appropriate teacher decoders, pooled into a global label space, used to train classical and VQC meta-decoders, and evaluated using teacher-label accuracy, confidence-gated fallback, and logical-failure analysis.}
\label{fig:qec_methodology_overview}
\end{figure*}

This section describes the proposed learned-assisted meta-decoding pipeline, shown in Fig.~\ref{fig:qec_methodology_overview}. The framework consists of six main components: problem formulation, code and noise specification, pooled dataset construction, meta-decoder training, hardware-aware VQC architecture search, and logical-level evaluation with confidence-gated recovery.

\subsection{Problem formulation}

\begin{figure*}[t]
\centering
\includegraphics[width=0.96\textwidth]{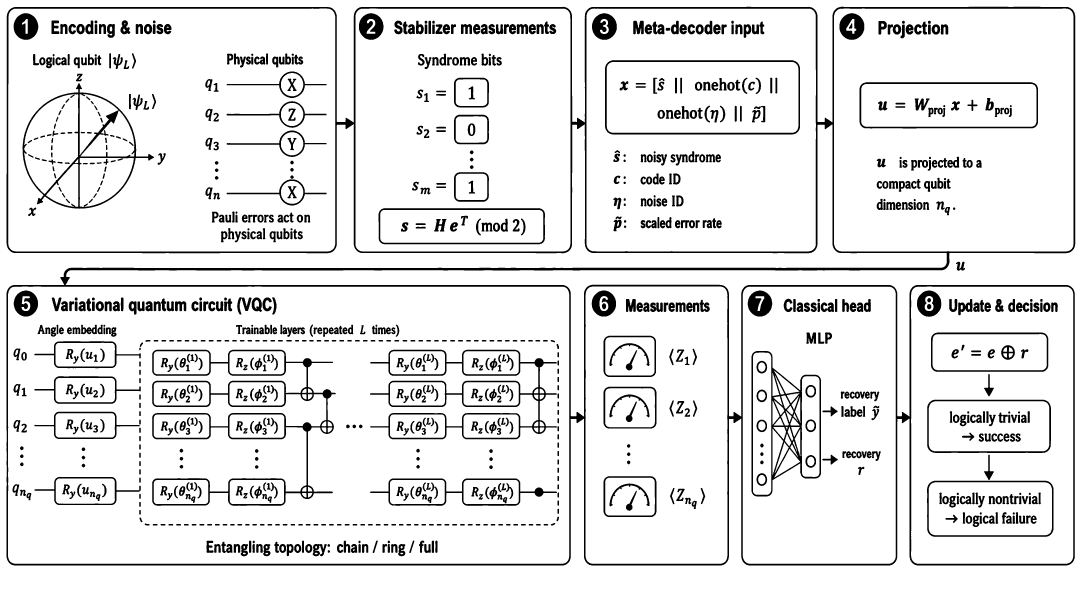}
\caption{Qubit-level mechanism of learned-assisted QEC meta-decoding. A logical qubit $|\psi_L\rangle$ is encoded into physical qubits, where Pauli errors generate stabilizer syndromes according to $s = H e^{T} \pmod{2}$. The syndrome, code identity, noise identity, and normalized physical error rate are combined into the meta-feature vector $x = [\tilde{s} \Vert \mathrm{onehot}(c) \Vert \mathrm{onehot}(\eta) \Vert \tilde{p}]$ and projected to a compact VQC input. The VQC applies angle embedding, trainable rotation layers, and entangling operations, after which Pauli-$Z$ measurements are mapped by a classical head to a recovery label and recovery operator $r$. Decoding success is determined by whether the corrected operator $e' = e \oplus r$ is logically trivial or logically nontrivial.}
\label{fig:qec_mechanism}
\end{figure*}

In a stabilizer-code setting, a physical Pauli error $e$ acting on $n$ qubits induces a syndrome through the binary symplectic product
\begin{equation}
s = H e^{T} \pmod 2,
\label{eq:method_syndrome}
\end{equation}
where $H$ is the binary stabilizer-check matrix of the code \cite{Gottesman1997Stabilizer}. Given syndrome $s$, the decoder must infer a recovery $r$ such that the corrected operator
\begin{equation}
e' = e \oplus r,
\label{eq:method_corrected_operator}
\end{equation}
does not induce a logical fault. If \(e'\) remains nontrivial at the logical level, decoding fails. The corresponding qubit-level mechanism, from physical Pauli errors and stabilizer syndrome extraction to learned recovery prediction and logical decision, is illustrated in Fig.~\ref{fig:qec_mechanism}.

We formulate this task as meta-decoding. Rather than training a separate decoder for each code and noise condition, a single model is trained on tuples
\begin{equation}
(s,c,\eta,p)\mapsto y,
\label{eq:method_meta_task}
\end{equation}
where $c$ is the code identifier, $\eta$ is the noise-family identifier, $p$ is the physical error probability, and $y$ is a discrete recovery-class label. The learned decoder is therefore represented as
\begin{equation}
f_{\theta}(s,c,\eta,p)=\hat{y},
\label{eq:method_meta_decoder}
\end{equation}
where $\theta$ denotes trainable parameters and $\hat{y}$ is mapped back to a recovery operator for logical evaluation.

The supervised objective is teacher-label classification: the model is trained to predict recovery labels produced by a code-appropriate teacher decoder. Teacher-label accuracy measures how well the learned model imitates this reference decoder, but it is not identical to the operational QEC objective because physically distinct recovery strings can be logically equivalent. For this reason, the evaluation distinguishes teacher-label classification accuracy from logical-level performance. Teacher-label accuracy measures supervised imitation and transfer, whereas logical-level performance is estimated through logical failure after applying the predicted or hybrid recovery.

\subsection{Meta-dataset construction}
The meta-dataset is constructed from four QEC settings: the compact algebraic FiveQubit \([[5,1,3]]\) stabilizer code, the Steane \([[7,1,3]]\) code, and two planar topological-code instances, Planar3$\times$3 and Planar5$\times$5. Teacher decoders are selected according to code family. This code-dependent choice reflects the structure of the benchmark. FiveQubit and Steane have compact syndrome spaces, so \texttt{NaiveDecoder} provides a consistent reference decoder for supervised label generation. Planar3$\times$3 and Planar5$\times$5 are planar topological-code instances, for which matching-based decoding is aligned with the lattice structure.

To evaluate transfer across heterogeneous physical conditions, samples are generated under four noise families: depolarizing noise, (Z)-biased Pauli noise, correlated (Z)-burst noise, and depolarizing noise with syndrome-measurement flips. For the (Z)-biased model, the single-qubit Pauli distribution is
\begin{equation}
\begin{aligned}
\Pr(I)&=1-p, & \Pr(X)&=0.2p,\\
\Pr(Y)&=0.1p, & \Pr(Z)&=0.7p.
\end{aligned}
\label{eq:method_zbiased}
\end{equation}
The coefficients $(0.2,0.1,0.7)$ are not intended as universal hardware-calibrated constants; rather, they define a representative moderately $Z$-biased Pauli channel in which the total non-identity error probability remains $p$, while $Z$-type errors dominate.
The correlated (Z)-burst model augments local Pauli errors with adjacent correlated (Z)-type events, while the measurement-flip variant additionally flips syndrome bits with probability proportional to the physical error rate. Supervised data are generated at
\begin{equation}
p \in \{0.03,0.05,0.07\},
\label{eq:method_p_values}
\end{equation}
with logical-failure curves evaluated later on a denser physical-error grid.

For each code, noise family, and physical error rate, raw syndrome-recovery records are first generated as
\begin{equation}
\bigl(s_i, c_i, \eta_i, p_i, r_i\bigr),
\label{eq:method_raw_tuple}
\end{equation}
where $s_i$ is the syndrome, $c_i$ is the code identifier, $\eta_i$ is the noise-family identifier, $p_i$ is the physical error probability, and $r_i$ is the teacher recovery. Because the codes have different numbers of stabilizers and physical qubits, all syndrome vectors and binary symplectic recovery strings are padded to common global dimensions:
\begin{equation}
\tilde{s}_i \in \{0,1\}^{S_{\max}}, \qquad
\tilde{r}_i \in \{0,1\}^{2Q_{\max}}.
\label{eq:method_padding}
\end{equation}

All raw records are pooled before the final train/test split. Each distinct code-specific padded recovery string is then assigned a global recovery label,
\begin{equation}
y_i = \mathrm{LabelMap}(c_i,\tilde{r}_i).
\label{eq:method_labelmap}
\end{equation}
Including the code identifier in the label key preserves code-specific recovery semantics while allowing all models to train over one unified supervised label space.

For non-held-out settings, the pooled dataset is split using a label-aware protocol. Singleton labels are assigned to training, labels with multiple occurrences retain at least one representative in training, and test labels are restricted to labels already represented in the training split. This prevents interpolation test classes from being absent during training. For held-out transfer settings, unseen test labels are allowed when they arise from the transfer design, since those settings intentionally evaluate out-of-distribution generalization.

The resulting benchmark includes one interpolation setting and four transfer settings: unseen-$p$ transfer, unseen-noise transfer, few-shot unseen-code adaptation, and few-shot held-out-size adaptation. In the main unseen-$p$ experiment, \(p=0.05\) is withheld during training. In the unseen-noise experiment, the (Z)-biased noise model is held out. In the few-shot unseen-code experiment, Planar5$\times$5 is held out from full training and only 5\% of its records are included as support. In the held-out-size experiment, code size (5) is held out from full training with 5\% support. Table~\ref{tab:split_summary} summarizes the resulting split sizes and unseen-label counts. The sample counts reported are the resulting train-test counts obtained after applying the dataset-generation and split protocols, rather than independently tuned hyperparameters.

\begin{table}[t]
\centering
\caption{Dataset split summary for the five evaluation settings. The global recovery-label space contains 15,582 code-specific recovery labels. Unseen test labels denote labels present in the test split but absent from the training split.}
\label{tab:split_summary}
\small
\begin{adjustbox}{max width=\columnwidth}
\begin{tabular}{lrrr}
\toprule
Setting & Train samples & Test samples & Unseen test labels \\
\midrule
Interpolation & 147,762 & 22,746 & 0 \\
Unseen-$p$ transfer & 99,726 & 54,758 & 4,515 \\
Unseen-noise transfer & 111,264 & 46,912 & 2,920 \\
Few-shot unseen-code adaptation & 76,500 & 46,500 & 13,529 \\
Few-shot held-out-size adaptation & 54,000 & 69,000 & 13,505 \\
\bottomrule
\end{tabular}
\end{adjustbox}
\end{table}
\subsection{Meta-decoder models and training protocol}

Each sample is represented by concatenating the padded syndrome vector, a one-hot code identifier, a one-hot noise-family identifier, and the normalized physical error rate. The resulting meta-feature vector is
\begin{equation}
x = [\,\tilde{s}\,\|\,\mathrm{onehot}(c)\,\|\,\mathrm{onehot}(\eta)\,\|\,\tilde{p}\,]
\label{eq:method_feature_vector}
\end{equation}
where \(\eta\) denotes the noise-family identifier and
\begin{equation}
\tilde{p} = \frac{p}{0.1}.
\label{eq:method_normalized_p}
\end{equation}
This representation allows one shared decoder to condition its prediction on the syndrome, code family, noise family, and physical error rate.

The classical baseline is a meta-decoder based on a multilayer perceptron, denoted Meta-MLP. Given input vector \(x\), the forward map is
\begin{align}
h_1 &= \phi(W_1x+b_1), \\
h_2 &= \phi(W_2h_1+b_2), \\
z &= W_3h_2+b_3,
\label{eq:method_mlp_forward}
\end{align}
where \(\phi(\cdot)\) is the ReLU activation and \(z\) contains logits over the global recovery-label space. The implementation uses two hidden layers with layer normalization and dropout.

To account for the greater difficulty of the larger planar code, the training protocol applies Planar5$\times$5 oversampling and a sample-weighted cross-entropy loss. The weighted objective is
\begin{equation}
\mathcal{L}_{\mathrm{wCE}}
=
\frac{1}{N}
\sum_{i=1}^{N}
w_i\,
\ell_{\mathrm{CE}}(z_i,y_i),
\label{eq:method_weighted_ce}
\end{equation}
where \(w_i\) increases the contribution of harder training regimes. In the final implementation, Planar5$\times$5 samples are oversampled by a factor of 3. The sample weight is factorized into code and noise components: the code weights are 1.0, 1.0, 1.2, and 3.0 for FiveQubit, Steane, Planar3$\times$3, and Planar5$\times$5, respectively, while the noise weights are 1.0, 1.0, 1.1, and 1.1 for depolarizing, (Z)-biased, correlated (Z)-burst, and depolarizing noise with syndrome-measurement flips, respectively. This weighting makes the supervised objective more sensitive to regimes that are harder at the logical level.

The quantum counterpart is a hybrid variational quantum classifier. Because the meta-feature vector can exceed the number of available circuit qubits, the input is first projected to the qubit dimension:
\begin{equation}
u = W_{\mathrm{proj}}x+b_{\mathrm{proj}},
\label{eq:method_proj}
\end{equation}
where \(u\in \mathbb{R}^{n_q}\) and \(n_q\) is the qubit count of the candidate VQC architecture. The projected vector \(u\) is embedded through angle embedding into a parameterized quantum circuit. Each circuit contains \(L\) trainable layers, with each layer applying single-qubit \(R_Y\) and \(R_Z\) rotations on every qubit followed by an entangling pattern selected from the architecture search space. The quantum feature vector is defined by Pauli-(Z) expectation values,
\begin{equation}
q(u)=\bigl[\langle Z_1\rangle,\ldots,\langle Z_{n_q}\rangle\bigr],
\label{eq:method_quantum_features}
\end{equation}
which are passed to a classical linear head:
\begin{equation}
\hat{z}=W_{\mathrm{head}}q(u)+b_{\mathrm{head}}.
\label{eq:method_quantum_head}
\end{equation}
The VQC meta-decoder is trained with the same weighted cross-entropy objective as the Meta-MLP.

Because recovery labels are code-specific, prediction is restricted using a code-label mask. For a sample from code \(c\), logits corresponding to labels associated with other codes are suppressed:
\begin{equation}
z_k \leftarrow -\infty
\qquad \text{for all } k \notin \mathcal{Y}(c),
\label{eq:method_code_mask}
\end{equation}
where \(\mathcal{Y}(c)\) is the set of valid labels for code \(c\). This masking step prevents cross-code recovery-label predictions and ensures that the global label space remains consistent with code-specific recovery semantics.
\subsection{Hardware-aware VQC architecture search}

The VQC architecture is selected through a compact hardware-aware quantum architecture search rather than fixed manually. The search space varies three architectural factors:
\begin{itemize}
\item qubit count $n_q \in \{4,5,6\}$,
\item circuit depth $L \in \{1,2\}$,
\item entangling pattern $\mathcal{E}\in\{\text{chain},\text{ring},\text{full}\}$.
\end{itemize}
This defines a discrete search space of 18 candidate architectures.

The search procedure has two stages. In the search stage, each candidate architecture is trained for 10 epochs and evaluated using validation accuracy. In the retraining stage, the architecture selected by validation accuracy and the architecture selected by hardware-aware score are each retrained from scratch for 35 epochs. This protocol keeps the search computationally manageable while ensuring that the final selected VQC models are evaluated after the same full retraining budget. The search is therefore intended as a compact NISQ-oriented architecture-selection study rather than a large combinatorial search.

To account for implementation cost, each architecture is also evaluated using hardware-oriented proxies, including raw circuit depth, raw two-qubit gate count, transpiled depth, and transpiled two-qubit gate count. For a candidate architecture $a$, the hardware-aware score is defined as
\begin{equation}
\begin{aligned}
S(a)={}&\mathrm{Acc}_{\mathrm{val}}(a)
-\lambda_1 D_{\mathrm{raw}}(a)
-\lambda_2 G^{\mathrm{raw}}_{2q}(a)\\
&-\lambda_3 D_{\mathrm{tr}}(a)
-\lambda_4 G^{\mathrm{tr}}_{2q}(a).
\end{aligned}
\label{eq:method_hardware_score}
\end{equation}
where $D$ denotes circuit depth and $G_{2q}$ denotes two-qubit gate count, each measured in raw and transpiled forms. The coefficients $\lambda_1,\lambda_2,\lambda_3,\lambda_4$ control the penalty assigned to circuit cost. The best-by-hardware-score VQC is selected according to $S(a)$, whereas the best-by-accuracy VQC is selected using validation accuracy alone.

\subsection{Confidence-gated recovery and logical evaluation}
The larger planar code, Planar5$\times$5, is especially sensitive to raw learned-decoder errors at the logical level. To reduce this risk, we use a confidence-gated fallback mechanism. Given the masked output distribution of a learned decoder, the prediction confidence is defined as
\begin{equation}
c(x)=\max_y p_{\theta}(y\mid x).
\label{eq:method_confidence}
\end{equation}
The final recovery rule is
\begin{equation}
r(x)=
\begin{cases}
r_{\mathrm{learned}}(x), & c(x)\ge \tau, \\
r_{\mathrm{teacher}}(x), & c(x)<\tau,
\end{cases}
\label{eq:method_fallback_rule}
\end{equation}
where $r_{\mathrm{teacher}}$ is the code-appropriate teacher recovery and $\tau$ is a decoder-specific confidence threshold. In this study, fallback is enabled only for Planar5$\times$5. The fixed thresholds are $\tau=0.80$ for the Meta-MLP and $\tau=0.75$ for the VQC decoders. These thresholds are treated as fixed operating points.

Logical-level evaluation is performed through Monte Carlo simulation. For each code, noise family, decoder, and physical error rate, we estimate
\begin{equation}
p_L(p)=\frac{N_{\mathrm{fail}}}{N_{\mathrm{run}}},
\label{eq:method_pl_estimate}
\end{equation}
where $N_{\mathrm{run}}$ is the number of sampled error instances and $N_{\mathrm{fail}}$ is the number of cases in which the corrected operator remains logically nontrivial. To compare learned decoders against the teacher, we use the logical-failure ratio
\begin{equation}
R_{\mathrm{logic}}(p)=
\frac{p_L^{(\mathrm{learned})}(p)}
{p_L^{(\mathrm{teacher})}(p)}.
\label{eq:method_logic_ratio}
\end{equation}
Values near $1$ indicate teacher-level logical behavior. In the final benchmarking, logical-failure curves are evaluated on the grid $p\in\{0.01,0.02,\ldots,0.10\}$ using 20,000 Monte Carlo trials per point.

\section{Results and Discussion}
\label{sec:results}
\subsection{Experimental setup}
\label{subsec:experimental_setup}

The full pipeline is implemented in Python using \texttt{qecsim} for code simulation and teacher decoding, \texttt{PyTorch} for the classical Meta-MLP, \texttt{PennyLane} for the VQC meta-decoders, and \texttt{Qiskit} for hardware-aware circuit costing and export. Table~\ref{tab:experimental_setup} summarizes the experimental settings and implementation details used in this study.

\begin{table*}[t]
\centering
\caption{Experimental setup and implementation details used in this study.}
\label{tab:experimental_setup}
\small
\begin{tabular}{p{0.30\textwidth}p{0.64\textwidth}}
\toprule
Parameter/Function & Value \\
\midrule
Physical error rates for supervised data & $p\in\{0.03,0.05,0.07\}$ \\
Logical-curve grid & $p\in\{0.01,0.02,\ldots,0.10\}$ \\
Logical Monte Carlo trials & 20,000 per point \\
Global syndrome padding & 40 stabilizer bits \\
Global recovery padding & 82 binary symplectic bits \\
Global label-space size & 15,582 labels \\
Meta features & syndrome, code one-hot, noise one-hot, normalized $p/0.1$ \\
MLP hidden dimensions & 512, 256 \\
MLP activation/normalization/dropout & ReLU, LayerNorm, dropout 0.15 \\
MLP optimizer & Adam, learning rate $10^{-3}$, weight decay $10^{-5}$ \\
MLP batch size / epochs & 256 / 35 \\
Planar5$\times$5 oversampling factor & 3 \\
Code weights & FiveQubit 1.0; Steane 1.0; Planar3$\times$3 1.2; Planar5$\times$5 3.0 \\
Noise weights & Depolarizing 1.0; $Z$-biased 1.0; correlated $Z$-burst 1.1; depol.+meas. flip 1.1 \\
VQC simulator & PennyLane \texttt{default.qubit} statevector simulator \\
VQC search space & $n_q\in\{4,5,6\}$, $L\in\{1,2\}$, entanglement in chain/ring/full \\
VQC optimizer & Adam, learning rate $10^{-2}$ \\
VQC batch size / epochs & 32; 10 search epochs and 35 final retraining epochs \\
Qiskit backend for cost proxy & \texttt{FakeManilaV2},  \texttt{FakeManila} \\
Transpilation optimization level & 1 \\
Fallback code family & Planar5$\times$5 only \\
Fallback thresholds & Meta-MLP $\tau=0.80$; VQC $\tau=0.75$ \\
\bottomrule
\end{tabular}
\end{table*}

\subsection{Teacher-label accuracy and VQC selection}

We first evaluate the pooled meta-decoding framework at the teacher-label level across five regimes: interpolation, unseen-$p$ transfer, unseen-noise transfer, few-shot unseen-code adaptation, and few-shot held-out-size adaptation. The evaluated methods are a majority-label baseline, the classical Meta-MLP, the VQC selected by search-stage validation accuracy, and the VQC selected by the hardware-aware score.

Table~\ref{tab:overall_summary} reports the main teacher-label accuracy results. The Meta-MLP is the strongest raw classifier in all five settings. It reaches 0.9993 in interpolation and remains strong under unseen-$p$ and unseen-noise transfer, with accuracies of 0.9118 and 0.9342, respectively. The largest drops occur in the few-shot adaptation regimes, where the accuracy decreases to 0.6304 for unseen-code adaptation and 0.7548 for held-out-size adaptation. This indicates that shifts in physical error rate or noise family are easier to absorb than changes in code structure or code size.

The VQC meta-decoders follow the same difficulty pattern but remain below the Meta-MLP in raw teacher-label accuracy. Still, both VQC variants stay well above the majority baseline in every regime, showing that they learn nontrivial syndrome-to-recovery structure. The hardware-aware VQC also remains close to the accuracy-selected VQC after retraining and gives higher accuracy in four of the five settings. The only exception is few-shot unseen-code adaptation, where the accuracy-selected VQC is slightly higher. Thus, the VQC results are best interpreted as compact quantum meta-decoders selected under different criteria, rather than as accuracy winners over the classical Meta-MLP.

\begin{table}[t]
\centering
\caption{Overall teacher-label classification accuracy across the five evaluation settings.}
\label{tab:overall_summary}
\begin{adjustbox}{max width=\columnwidth}
\begin{tabular}{lcccc}
\toprule
Setting & Majority & Meta-MLP & VQC best-acc & VQC best-hw \\
\midrule
Interpolation & 0.2032 & 0.9993 & 0.8987 & 0.9400 \\
unseen-$p$ transfer & 0.1966 & 0.9118 & 0.8397 & 0.8495 \\
Unseen-noise transfer & 0.1975 & 0.9342 & 0.8384 & 0.8415 \\
Few-shot unseen-code adaptation & 0.0994 & 0.6304 & 0.5435 & 0.5678 \\
Few-shot held-out-size adaptation & 0.0599 & 0.7548 & 0.6774 & 0.7143 \\
\bottomrule
\end{tabular}
\end{adjustbox}
\end{table}

\begin{figure}[t]
\centering
\includegraphics[width=0.98\linewidth]{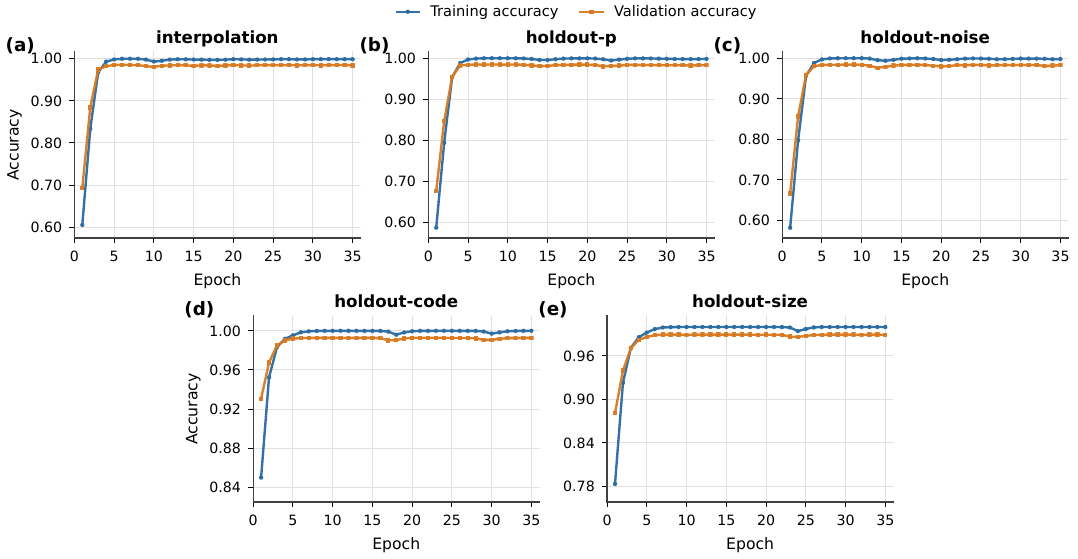}
\caption{Training and validation accuracy of the Meta-MLP on the pooled multi-code, multi-noise dataset.}
\label{fig:mlp-training-curves}
\end{figure}

\begin{figure}[t]
\centering
\includegraphics[width=0.98\linewidth]{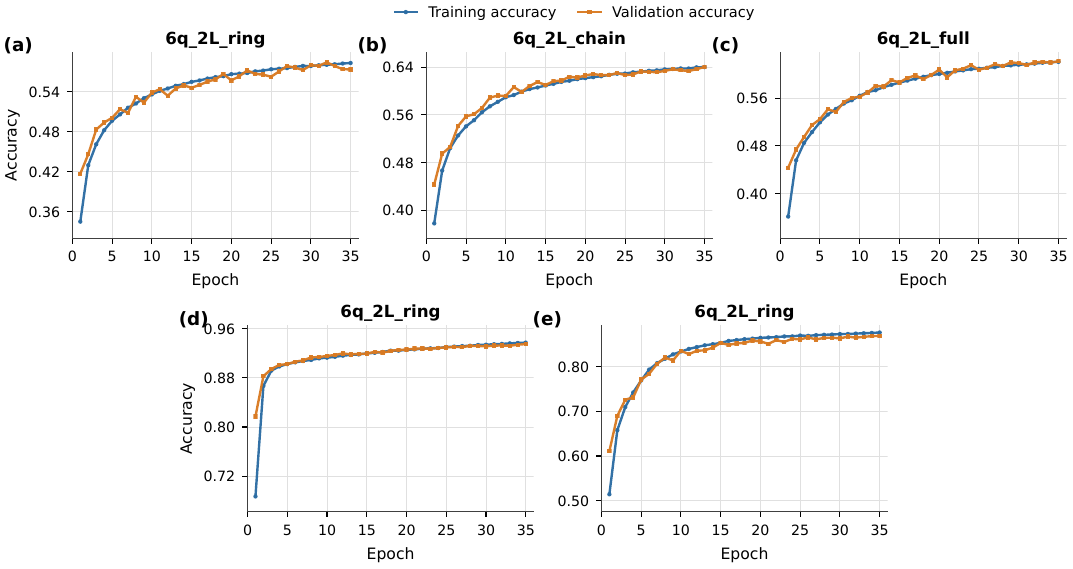}
\caption{Retraining curves of the two selected VQC meta-decoders. Panels (a)--(e) correspond to interpolation, unseen-$p$ transfer, unseen-noise transfer, few-shot unseen-code adaptation, and few-shot held-out-size adaptation, respectively.}
\label{fig:vqc-retain-curves}
\end{figure}

The learning curves support this interpretation. The Meta-MLP converges rapidly and stably on the pooled syndrome-plus-metadata representation, as shown in Fig.~\ref{fig:mlp-training-curves}. This behavior is consistent with the near-saturated interpolation result in Table~\ref{tab:overall_summary}. In contrast, the selected VQCs converge more slowly and reach lower final accuracy, as shown in Fig.~\ref{fig:vqc-retain-curves}. However, both VQC variants improve during retraining across all five regimes, including the few-shot transfer settings.

\begin{figure}[t]
\centering
\includegraphics[width=0.98\linewidth]{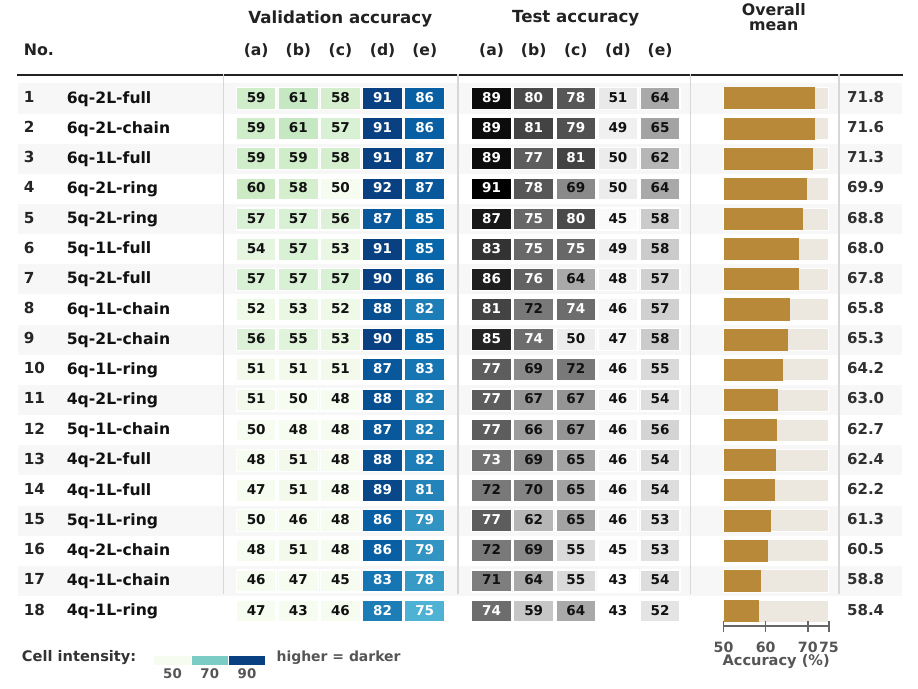}
\caption{Validation and search-stage test accuracy of candidate VQC architectures explored during quantum architecture search.}
\label{fig:qas-acc}
\end{figure}

\begin{figure}[t]
\centering
\includegraphics[width=0.98\linewidth]{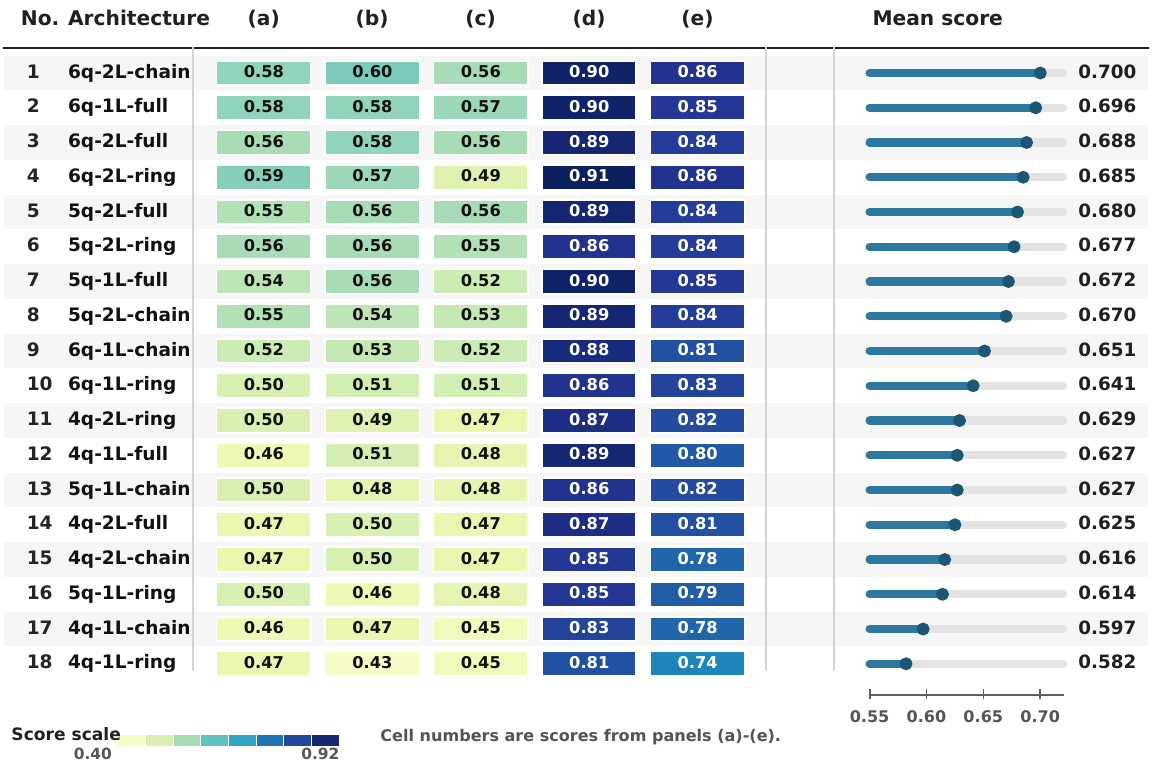}
\caption{Hardware-aware score of candidate VQC architectures. The score combines predictive performance with implementation-oriented circuit-cost proxies such as depth and two-qubit gate count.}
\label{fig:qas-hwscore}
\end{figure}

The VQC architecture-search results show that the candidate circuits are not equivalent. Fig.~\ref{fig:qas-acc} reports validation and search-stage test accuracy, while Fig.~\ref{fig:qas-hwscore} reports the hardware-aware score. Because the candidates differ in qubit count, circuit depth, and entangling topology, the search produces distinct accuracy-cost tradeoffs. The hardware-aware criterion therefore acts as a model-selection rule during VQC search, not only as a post hoc circuit-cost analysis.

\begin{figure}[!htpb]
\centering
\begin{overpic}[width=\columnwidth,trim=8 8 8 8,clip]{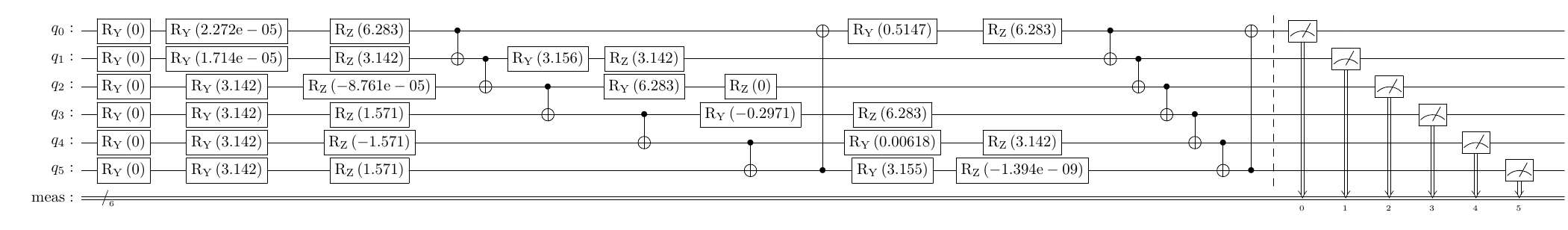}
\put(-1.2,2.5){\footnotesize \textbf{(a)}}
\end{overpic}
\vspace{-0.9em}

\begin{overpic}[width=\columnwidth,trim=8 8 8 8,clip]{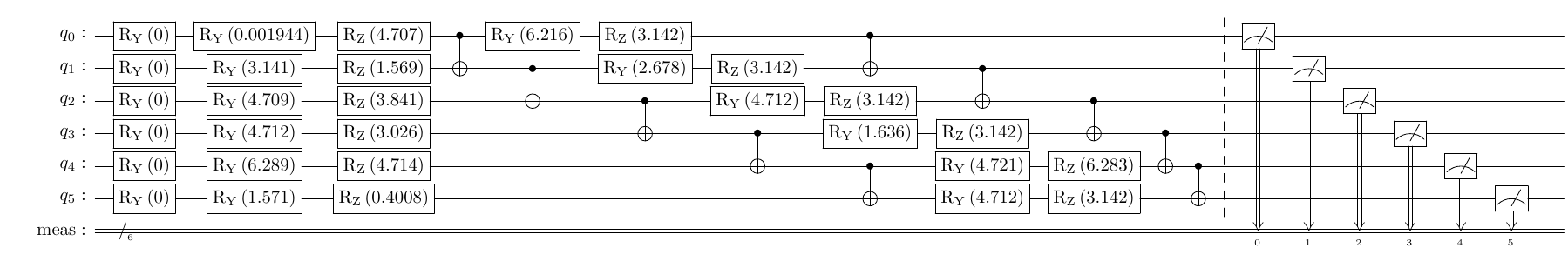}
\put(-1.2,2.5){\footnotesize \textbf{(b)}}
\end{overpic}
\vspace{-0.9em}

\begin{overpic}[width=\columnwidth,trim=8 8 8 8,clip]{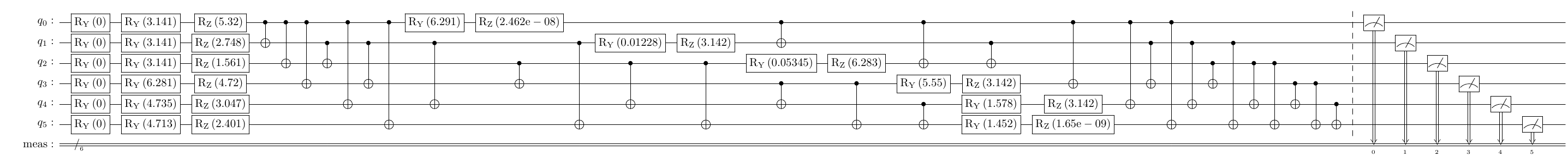}
\put(-1.2,2.5){\footnotesize \textbf{(c)}}
\end{overpic}
\vspace{-0.9em}

\begin{overpic}[width=\columnwidth,trim=8 8 8 8,clip]{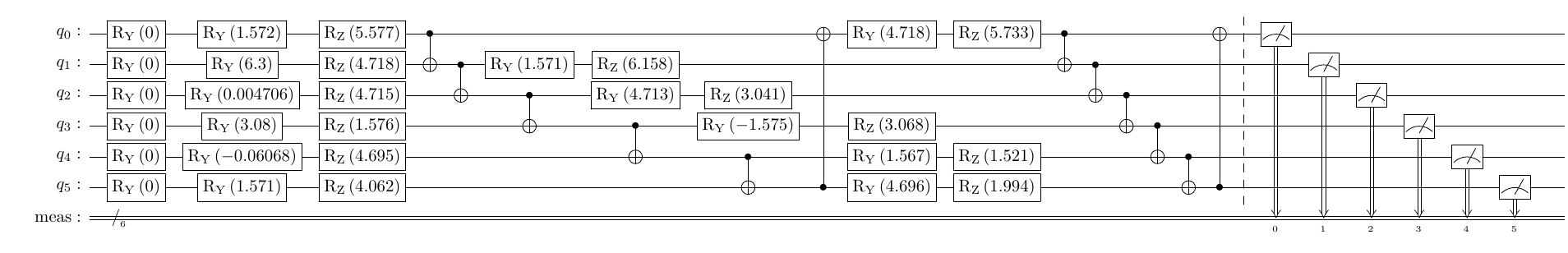}
\put(-1.2,2.5){\footnotesize \textbf{(d)}}
\end{overpic}
\vspace{-0.9em}

\begin{overpic}[width=\columnwidth,trim=8 8 8 8,clip]{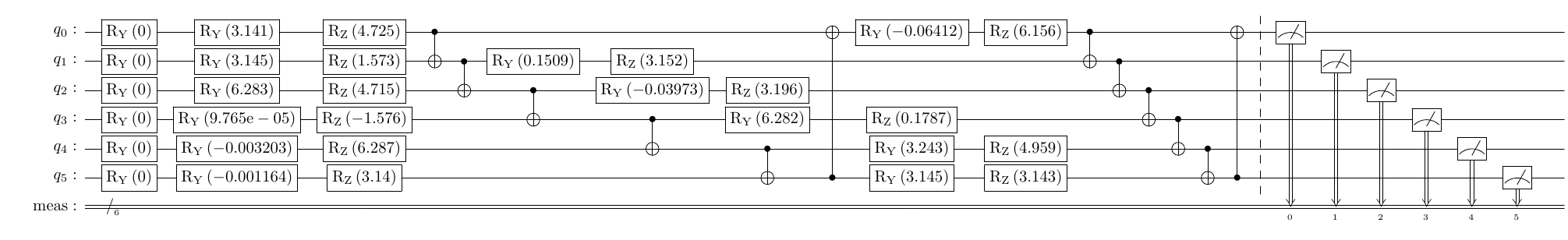}
\put(-1.2,2.5){\footnotesize \textbf{(e)}}
\end{overpic}
\caption{Selected VQC architectures obtained using search-stage validation accuracy. Panels (a)--(e) correspond to interpolation, unseen-$p$ transfer, unseen-noise transfer, few-shot unseen-code adaptation, and few-shot held-out-size adaptation, respectively.}
\label{fig:architecture_acc}
\end{figure}

\begin{figure}[htpbt]
\centering

\begin{overpic}[width=\columnwidth,trim=8 8 8 8,clip]{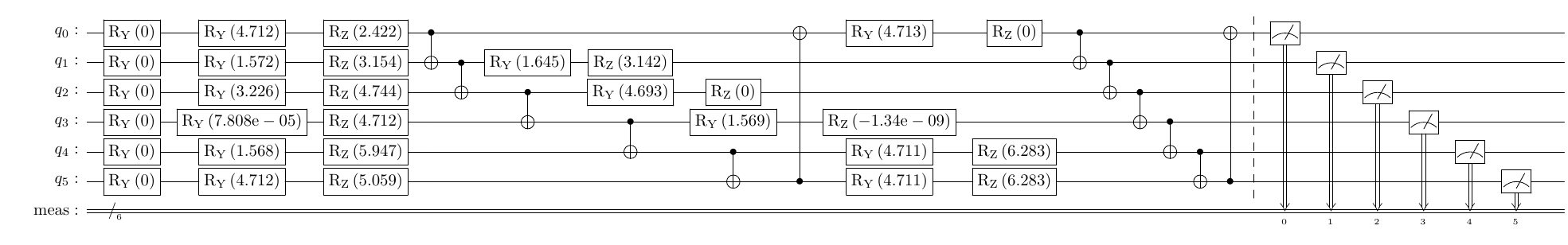}
\put(-1.2,2.5){\footnotesize \textbf{(a)}}
\end{overpic}
\vspace{-0.9em}

\begin{overpic}[width=\columnwidth,trim=8 8 8 8,clip]{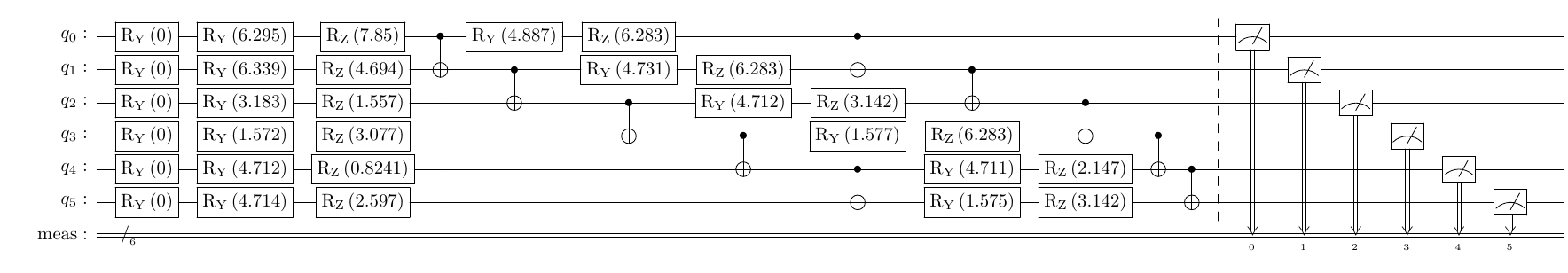}
\put(-1.2,2.5){\footnotesize \textbf{(b)}}
\end{overpic}
\vspace{-0.9em}

\begin{overpic}[width=\columnwidth,trim=8 8 8 8,clip]{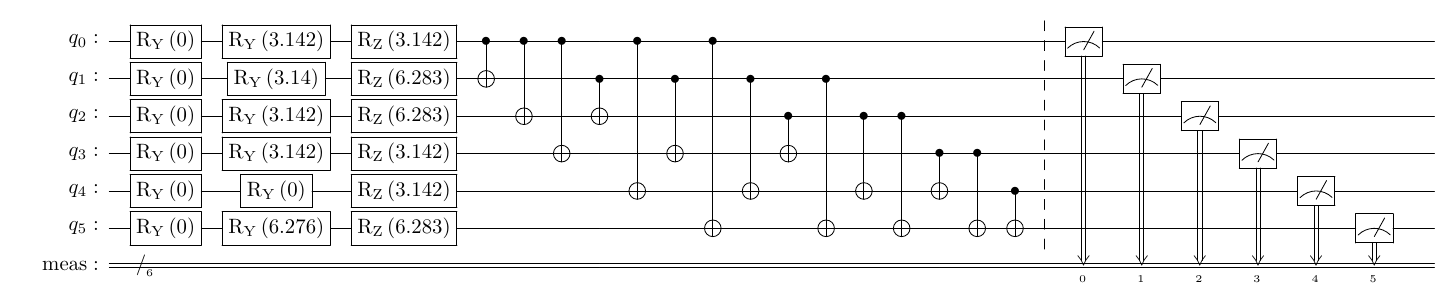}
\put(-1.2,2.5){\footnotesize \textbf{(c)}}
\end{overpic}
\vspace{-0.9em}

\begin{overpic}[width=\columnwidth,trim=8 8 8 8,clip]{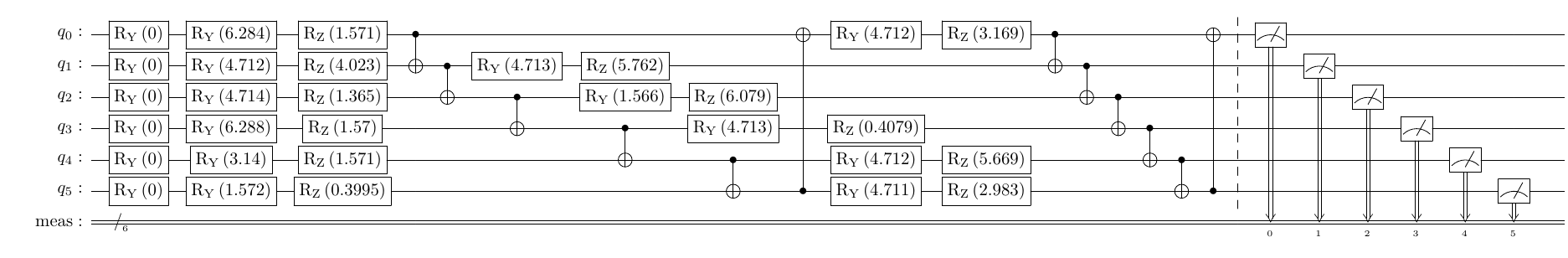}
\put(-1.2,2.5){\footnotesize \textbf{(d)}}
\end{overpic}
\vspace{-0.9em}

\begin{overpic}[width=\columnwidth,trim=8 8 8 8,clip]{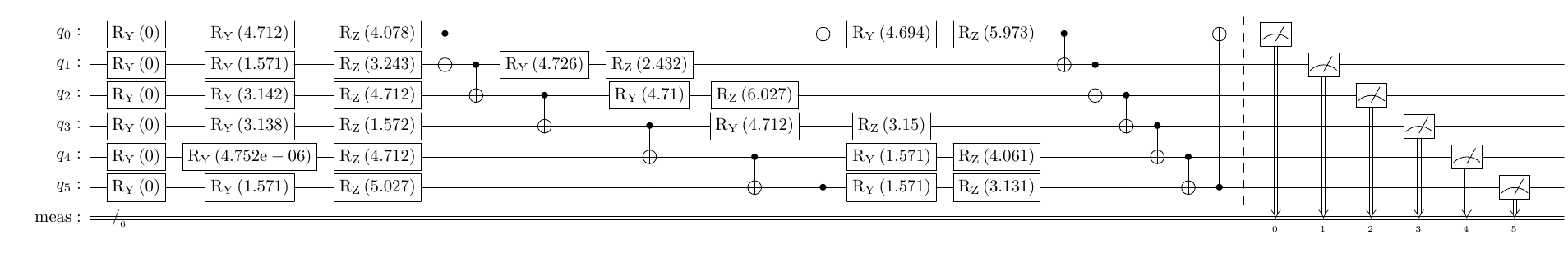}
\put(-1.2,2.5){\footnotesize \textbf{(e)}}
\end{overpic}
\caption{Selected VQC architectures obtained using the hardware-aware score. Panels (a)--(e) correspond to interpolation, unseen-$p$ transfer, unseen-noise transfer, few-shot unseen-code adaptation, and few-shot held-out-size adaptation, respectively.}
\label{fig:architecture_hardware}
\end{figure}

The selected circuit architectures in Fig.~\ref{fig:architecture_acc} and Fig.~\ref{fig:architecture_hardware} show the effect of the two selection criteria at the circuit level. Compared with the accuracy-selected circuit, the hardware-aware selection chooses a lower-cost architecture while retaining competitive teacher-label accuracy. 

\subsection{Interpolation}
\label{subsec:interpolation}

We begin with the interpolation setting, where all code families, noise models, and supervised physical error rates are represented during training. This setting tests whether the pooled meta-decoder can learn a shared syndrome-to-recovery map in the absence of distributional shift.

As shown in Table~\ref{tab:overall_summary}, the Meta-MLP achieves a test accuracy of 0.9993, far above the majority baseline of 0.2032. The VQC models also perform well above the baseline, with accuracies of 0.8987 for the accuracy-selected VQC and 0.9400 for the hardware-aware VQC. In this setting, the hardware-aware VQC outperforms the accuracy-selected VQC after retraining, showing that cost-aware selection does not necessarily reduce teacher-label accuracy within the considered VQC search space.

The code-wise interpolation results in Table~\ref{tab:interp_codewise} show that the algebraic codes are the easiest cases. FiveQubit is solved by all learned models, and Steane is nearly saturated by the Meta-MLP while remaining high for both VQCs. The planar codes are more difficult for the VQC models. Planar3$\times$3 remains manageable, but Planar5$\times$5 is the main bottleneck: the Meta-MLP reaches 0.999, whereas the accuracy-selected and hardware-aware VQCs reach 0.692 and 0.830, respectively.

\begin{table}[t]
\centering
\caption{Average code-wise interpolation accuracy.}
\label{tab:interp_codewise}
\begin{adjustbox}{max width=\columnwidth}
\begin{tabular}{lccc}
\toprule
Code & Meta-MLP & VQC best-acc & VQC best-hw \\
\midrule
FiveQubit & 1.000 & 1.000 & 1.000 \\
Steane & 1.000 & 0.947 & 0.976 \\
Planar3$\times$3 & 0.998 & 0.870 & 0.887 \\
Planar5$\times$5 & 0.999 & 0.692 & 0.830 \\
\bottomrule
\end{tabular}
\end{adjustbox}
\end{table}

This code-wise pattern is consistent with the broader groupwise results in Fig.~\ref{fig:groupwise-acc}(a). Interpolation accuracy is nearly saturated for the Meta-MLP across most code-noise-\(p\) groups, while the VQC errors are concentrated in the planar-code settings. Thus, interpolation establishes the main in-distribution result: the pooled decoder can learn the shared teacher-label task, but the larger planar code remains the most difficult case for the compact VQC models.

Although interpolation is almost solved at the teacher-label level by the Meta-MLP, label accuracy alone does not fully determine decoder reliability. We therefore return to logical-failure behavior in Sec.~\ref{subsec:logical_fallback}, where Planar5$\times$5 is analyzed under the confidence-gated fallback protocol.

\subsection{Transfer to unseen physical error rate}
\label{subsec:unseen_p}

The unseen-$p$ setting tests whether the pooled representation generalizes across physical error magnitude rather than memorizing the specific \(p\)-values used in training. As shown in Table~\ref{tab:overall_summary}, all learned models remain clearly above the majority baseline in this regime. The Meta-MLP achieves 0.9118, while the accuracy-selected and hardware-aware VQCs reach 0.8397 and 0.8495, respectively. Relative to interpolation, the performance drop is noticeable but moderate, suggesting that the learned representation captures structure that persists across nearby physical error rates.

\begin{figure}[t]
\centering
\includegraphics[width=0.98\linewidth]{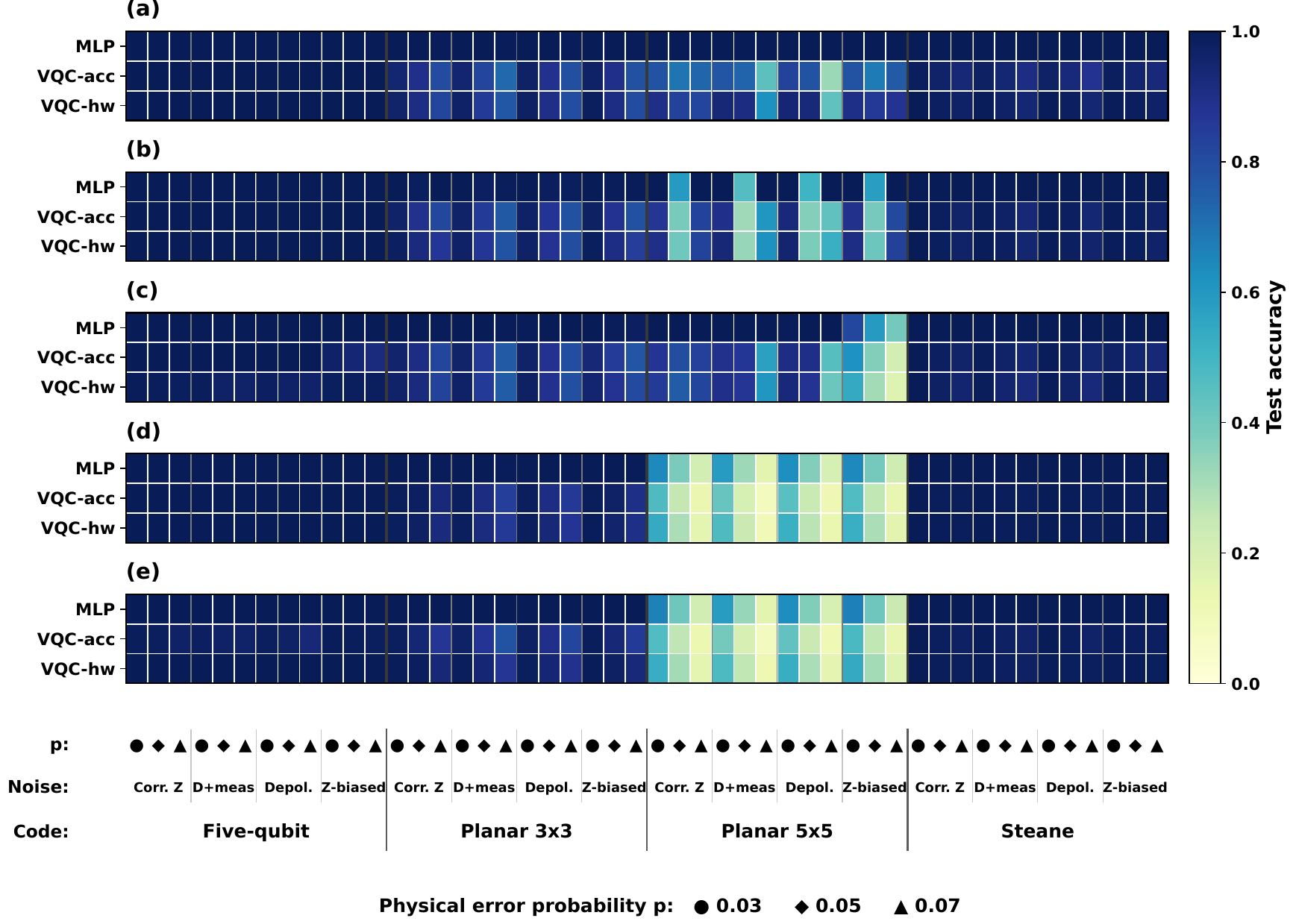}
\caption{Groupwise teacher-label accuracy across code, noise family, and physical error rate for the final benchmarked decoders. Panels (a)--(e) correspond to interpolation, unseen-$p$ transfer, unseen-noise transfer, few-shot unseen-code adaptation, and few-shot held-out-size adaptation, respectively.}
\label{fig:groupwise-acc}
\end{figure}

The code-wise decomposition in Table~\ref{tab:holdoutp_codewise} shows that the degradation is not uniform across codes. FiveQubit and Steane remain near-saturated, and Planar3$\times$3 stays close to 0.9 even for the VQCs. The main loss again occurs in Planar5$\times$5, where the Meta-MLP reaches 0.844, while the accuracy-selected and hardware-aware VQCs reach 0.644 and 0.670, respectively. This same progression is visible in Fig.~\ref{fig:groupwise-acc}(b): unseen-$p$ transfer remains robust for most groups, but the larger planar-code setting is where generalization degrades most clearly.

\begin{table}[t]
\centering
\caption{Average code-wise accuracy under unseen-$p$ transfer.}
\label{tab:holdoutp_codewise}
\begin{adjustbox}{max width=\columnwidth}
\begin{tabular}{lccc}
\toprule
Code & Meta-MLP & VQC best-acc & VQC best-hw \\
\midrule
FiveQubit & 1.000 & 1.000 & 1.000 \\
Steane & 1.000 & 0.974 & 0.978 \\
Planar3$\times$3 & 0.993 & 0.877 & 0.896 \\
Planar5$\times$5 & 0.844 & 0.644 & 0.670 \\
\bottomrule
\end{tabular}
\end{adjustbox}
\end{table}

Thus, unseen-$p$ transfer shows that changing the physical error magnitude is manageable at the teacher-label level, but Planar5$\times$5 remains the main bottleneck. 

\subsection{Transfer to unseen noise family}
\label{subsec:unseen_noise}

The unseen-noise setting is a stronger generalization test because it changes the structure of the corruption process rather than only its magnitude. As shown in Table~\ref{tab:overall_summary}, the pooled framework remains robust under this shift. The Meta-MLP achieves 0.9342, while the accuracy-selected and hardware-aware VQCs reach 0.8384 and 0.8415, respectively. These values are close to the unseen-$p$ setting, indicating that the syndrome-plus-metadata representation captures structure that transfers across noise families.

The code-wise results in Table~\ref{tab:holdoutnoise_codewise} again show that the difficulty is concentrated mainly in Planar5$\times$5. FiveQubit and Steane remain near-perfect, and Planar3$\times$3 stays strong. Planar5$\times$5 is harder, with accuracies of 0.900 for the Meta-MLP and 0.691 and 0.670 for the two VQCs. Fig.~\ref{fig:groupwise-acc}(c) shows the same pattern: the algebraic and smaller planar-code settings remain robust, while the larger planar code remains the main transfer bottleneck.

\begin{table}[t]
\centering
\caption{Average code-wise accuracy under unseen-noise transfer.}
\label{tab:holdoutnoise_codewise}
\begin{adjustbox}{max width=\columnwidth}
\begin{tabular}{lccc}
\toprule
Code & Meta-MLP & VQC best-acc & VQC best-hw \\
\midrule
FiveQubit & 1.000 & 0.986 & 0.980 \\
Steane & 1.000 & 0.967 & 0.968 \\
Planar3$\times$3 & 0.996 & 0.872 & 0.881 \\
Planar5$\times$5 & 0.900 & 0.691 & 0.670 \\
\bottomrule
\end{tabular}
\end{adjustbox}
\end{table}

Thus, unseen-noise transfer remains feasible at the teacher-label level, but the result also confirms that changing the noise family does not remove the structural difficulty of Planar5$\times$5. 

\subsection{Few-shot adaptation to an unseen code}
\label{subsec:fewshot_unseen_code}

The few-shot unseen-code setting is the most demanding transfer regime because the model must adapt to a held-out code configuration using only a small support fraction. In the present setup, the held-out code is Planar5$\times$5 with 5\% support. As shown in Table~\ref{tab:overall_summary}, this setting gives the strongest overall degradation among all experiments. The Meta-MLP reaches 0.6304, while the accuracy-selected and hardware-aware VQCs reach 0.5435 and 0.5678, respectively. Although these values remain above the majority baseline of 0.0994, they are far below the corresponding interpolation, unseen-$p$, and unseen-noise results.

Table~\ref{tab:holdoutcode_codewise} clarifies the source of this degradation. The non-held-out codes remain almost saturated, but Planar5$\times$5 drops sharply to 0.397 for the Meta-MLP and to 0.270 and 0.308 for the two VQCs. Fig.~\ref{fig:groupwise-acc}(d) shows that this drop is highly localized: most configurations remain easy, while the truly unseen code structure is substantially harder. This result shows that the pooled decoder transfers well across known settings, but still faces a clear representation gap when the syndrome-recovery map changes in a structurally new way.

\begin{table}[t]
\centering
\caption{Average code-wise accuracy under few-shot unseen-code adaptation.}
\label{tab:holdoutcode_codewise}
\begin{adjustbox}{max width=\columnwidth}
\begin{tabular}{lccc}
\toprule
Code & Meta-MLP & VQC best-acc & VQC best-hw \\
\midrule
FiveQubit & 1.000 & 1.000 & 1.000 \\
Steane & 1.000 & 0.994 & 0.995 \\
Planar3$\times$3 & 0.999 & 0.938 & 0.940 \\
Planar5$\times$5 & 0.397 & 0.270 & 0.308 \\
\bottomrule
\end{tabular}
\end{adjustbox}
\end{table}

Thus, few-shot adaptation to an unseen code exposes the clearest limitation of raw teacher-label transfer. The model retains strong performance on known code structures, but direct adaptation to Planar5$\times$5 remains weak under limited support data. 

\subsection{Few-shot adaptation to held-out code size}
\label{subsec:fewshot_heldout_size}

The held-out-size setting addresses a related but distinct question: whether the meta-decoder can adapt to a new code scale using only a small support set. As shown in Table~\ref{tab:overall_summary}, this regime is harder than unseen-$p$ and unseen-noise transfer, but easier than few-shot unseen-code adaptation. The Meta-MLP reaches 0.7548, while the accuracy-selected and hardware-aware VQCs reach 0.6774 and 0.7143, respectively. The hardware-aware VQC again exceeds the accuracy-selected VQC in this setting, suggesting that the hardware-aware criterion can select compact architectures that remain competitive under transfer.

The code-wise breakdown in Table~\ref{tab:holdoutsize_codewise} shows that the held-out-size result combines easy and difficult cases. FiveQubit, Steane, and Planar3$\times$3 remain strong, while Planar5$\times$5 remains the main bottleneck. In particular, the Meta-MLP reaches 0.406 on Planar5$\times$5, while the accuracy-selected and hardware-aware VQCs reach 0.266 and 0.321, respectively. This pattern is also visible in Fig.~\ref{fig:groupwise-acc}(e), where the larger planar-code setting remains the main source of difficulty.

\begin{table}[t]
\centering
\caption{Average code-wise accuracy under few-shot held-out-size adaptation.}
\label{tab:holdoutsize_codewise}
\begin{adjustbox}{max width=\columnwidth}
\begin{tabular}{lccc}
\toprule
Code & Meta-MLP & VQC best-acc & VQC best-hw \\
\midrule
FiveQubit & 1.000 & 0.976 & 1.000 \\
Steane & 1.000 & 0.983 & 0.989 \\
Planar3$\times$3 & 1.000 & 0.907 & 0.952 \\
Planar5$\times$5 & 0.406 & 0.266 & 0.321 \\
\bottomrule
\end{tabular}
\end{adjustbox}
\end{table}

Thus, held-out-size adaptation lies between unseen-noise transfer and unseen-code adaptation in difficulty. The results confirm that structural changes in code scale are harder than shifts in \(p\) or noise family, especially for the larger topological-code setting. 

\subsection{Logical-level reliability and selective fallback}
\label{subsec:logical_fallback}

Teacher-label accuracy is a useful first measure of meta-decoder performance, but it is not equivalent to logical reliability. In QEC decoding, a small number of incorrect recovery labels can still induce logical failures, particularly in larger topological-code settings. We therefore evaluate the learned decoders at the logical level and test whether confidence-gated fallback can reduce the gap between raw learned recovery and teacher-decoder behavior.

\begin{figure}[t]
\centering
\includegraphics[width=0.94\linewidth]{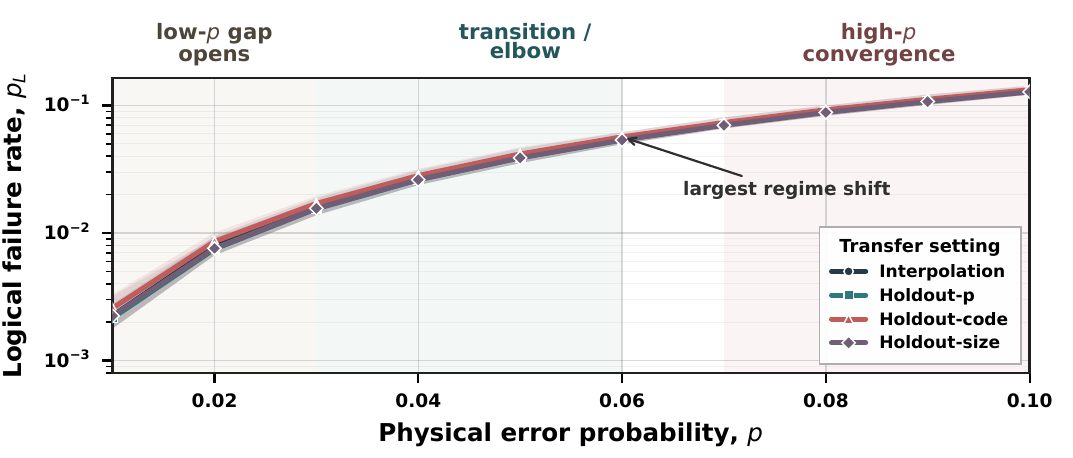}
\caption{Combined logical-failure curves under the confidence-gated hybrid recovery protocol.}
\label{fig:logical_failure}
\end{figure}

The logical-failure curves in Fig.~\ref{fig:logical_failure} show the resulting decoder behavior across the five evaluation settings. Planar5$\times$5 is the most demanding regime: raw learned decoding remains far from the teacher decoder, while confidence-gated fallback reduces the logical gap by using the learned decoder only on confident cases and routing uncertain cases to the teacher decoder (Additional Planar5$\times$5 curves are reported in Appendix~\ref{app:planar55_logical_curves}).

Table~\ref{tab:planar55_raw_fallback_all} quantifies the raw and fallback-enabled logical behavior on Planar5$\times$5. In interpolation, the Meta-MLP reaches nearly perfect teacher-label accuracy, yet its raw logical-failure ratio remains 12.08 relative to the teacher decoder. The raw VQC ratios are even larger, at 42.41 for the accuracy-selected VQC and 25.91 for the hardware-aware VQC. After fallback, these ratios drop to 1.71, 1.04, and 1.11, respectively. The same pattern appears across the transfer settings: raw learned decoding is not a reliable teacher replacement, while fallback brings the learned decoders much closer to teacher-level logical behavior.

\begin{table}[t]
\centering
\caption{Mean Planar5$\times$5 logical-failure ratio relative to the teacher decoder across the five settings. Values near 1 indicate teacher-level logical behavior.}
\label{tab:planar55_raw_fallback_all}
\scriptsize
\resizebox{\linewidth}{!}{%
\begin{tabular}{llrrr}
\toprule
Setting & Decoder & Raw ratio & Fallback ratio & Reduction factor \\
\midrule
Interpolation & Meta-MLP & 12.08 & 1.71 & 7.06 \\
Interpolation & VQC best-acc & 42.41 & 1.04 & 40.88 \\
Interpolation & VQC best-hw & 25.91 & 1.11 & 23.24 \\
unseen-$p$ & Meta-MLP & 14.61 & 1.92 & 7.61 \\
unseen-$p$ & VQC best-acc & 32.62 & 1.09 & 29.94 \\
unseen-$p$ & VQC best-hw & 35.15 & 1.06 & 33.03 \\
Unseen-noise & Meta-MLP & 14.52 & 1.77 & 8.19 \\
Unseen-noise & VQC best-acc & 40.39 & 1.04 & 38.83 \\
Unseen-noise & VQC best-hw & 43.27 & 3.17 & 13.66 \\
Few-shot unseen-code & Meta-MLP & 32.27 & 3.76 & 8.59 \\
Few-shot unseen-code & VQC best-acc & 65.80 & 1.09 & 60.25 \\
Few-shot unseen-code & VQC best-hw & 50.85 & 1.05 & 48.55 \\
Few-shot held-out-size & Meta-MLP & 30.76 & 3.20 & 9.63 \\
Few-shot held-out-size & VQC best-acc & 54.46 & 1.11 & 49.27 \\
Few-shot held-out-size & VQC best-hw & 58.13 & 1.09 & 53.23 \\
\bottomrule
\end{tabular}}
\end{table}

To make the fallback mechanism visible at the operating point used in the experiments, Table~\ref{tab:planar55_coverage_existing} reports fixed-threshold learned coverage and fallback rates on Planar5$\times$5. These quantities are computed from saved test-set confidence values, not from new simulations. They measure how often the learned decoder is used at the reported thresholds, while Table~\ref{tab:planar55_raw_fallback_all} reports the corresponding Monte Carlo logical behavior.

\begin{table*}[t]
\centering
\caption{Planar5$\times$5 selective-decoding coverage computed from the saved test-set confidence values. Meta-MLP uses \(\tau=0.80\) and VQC models use \(\tau=0.75\), matching the reported fallback protocol. Confident error is measured against the teacher recovery label, not against logical equivalence; it is therefore a teacher-label proxy rather than a logical-failure rate.}
\label{tab:planar55_coverage_existing}
\scriptsize
\resizebox{\textwidth}{!}{%
\begin{tabular}{llrrrrrr}
\toprule
Setting & Model & \(\tau\) & Coverage & Fallback & Label acc. & Conf. label acc. & Conf. label err. \\
\midrule
Interpolation & Meta-MLP & 0.80 & 0.9989 & 0.0011 & 0.9992 & 0.9998 & 0.0002 \\
Interpolation & VQC best-acc & 0.75 & 0.3335 & 0.6665 & 0.7463 & 0.9994 & 0.0006 \\
Interpolation & VQC best-hw & 0.75 & 0.6049 & 0.3951 & 0.8854 & 0.9986 & 0.0014 \\
Unseen-\(p\) & Meta-MLP & 0.80 & 0.6451 & 0.3549 & 0.6345 & 0.9673 & 0.0327 \\
Unseen-\(p\) & VQC best-acc & 0.75 & 0.2132 & 0.7868 & 0.4734 & 0.9941 & 0.0059 \\
Unseen-\(p\) & VQC best-hw & 0.75 & 0.3156 & 0.6844 & 0.4926 & 0.9923 & 0.0077 \\
Unseen-noise & Meta-MLP & 0.80 & 0.7361 & 0.2639 & 0.7257 & 0.9817 & 0.0183 \\
Unseen-noise & VQC best-acc & 0.75 & 0.2499 & 0.7501 & 0.5419 & 0.9982 & 0.0018 \\
Unseen-noise & VQC best-hw & 0.75 & 0.2431 & 0.7569 & 0.4997 & 0.9819 & 0.0181 \\
Few-shot unseen-code & Meta-MLP & 0.80 & 0.4204 & 0.5796 & 0.3971 & 0.9170 & 0.0830 \\
Few-shot unseen-code & VQC best-acc & 0.75 & 0.1585 & 0.8415 & 0.2697 & 0.9639 & 0.0361 \\
Few-shot unseen-code & VQC best-hw & 0.75 & 0.1629 & 0.8371 & 0.3085 & 0.9892 & 0.0108 \\
Few-shot held-out-size & Meta-MLP & 0.80 & 0.4093 & 0.5907 & 0.4064 & 0.9463 & 0.0537 \\
Few-shot held-out-size & VQC best-acc & 0.75 & 0.1561 & 0.8439 & 0.2660 & 0.9737 & 0.0263 \\
Few-shot held-out-size & VQC best-hw & 0.75 & 0.1616 & 0.8384 & 0.3208 & 0.9920 & 0.0080 \\
\bottomrule
\end{tabular}}
\end{table*}

The coverage results in Table~\ref{tab:planar55_coverage_existing} explain the source of this logical improvement. At \(\tau=0.75\), the VQC models are conservative on Planar5$\times$5 and often route most transfer cases to the teacher decoder. For example, in the few-shot regimes, their learned coverage stays near 0.16, while the confident label accuracy remains much higher than the full label accuracy. This indicates that the confidence signal is useful for selective decoding: the learned decoder is not used broadly, but the cases it retains are substantially more reliable.

\begin{figure}[t]
\centering
\includegraphics[width=0.98\linewidth]{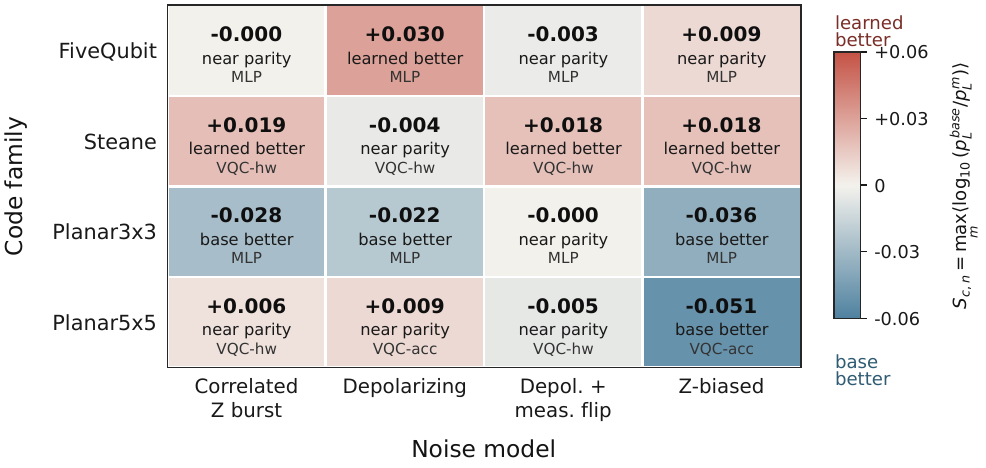}
\caption{Computed decoder-advantage summary across code families and noise models. Each cell reports the best learned-decoder advantage relative to the corresponding baseline decoder, averaged over transfer settings and physical error probabilities. Positive values indicate lower logical-failure rates for the best learned decoder, whereas negative values indicate that the baseline decoder remains superior.}
\label{fig:decoder_heatmap}
\end{figure}

Finally, Fig.~\ref{fig:decoder_heatmap} summarizes the learned-decoder advantage across code families and noise models. For each code--noise pair, the score is computed as
\begin{equation}
S_{c,n}=\max_m\left\langle
\log_{10}\left(\frac{p_L^{\mathrm{baseline}}}{p_L^{m}}\right)
\right\rangle_{t,p},
\end{equation}
where \(p_L^{\mathrm{baseline}}\) is the logical-failure rate of the code-specific conventional decoder, \(p_L^m\) is the logical-failure rate of the learned decoder \(m\), and the average is taken over transfer settings \(t\) and physical error probabilities \(p\). The baseline is the Naive decoder for FiveQubit and Steane, and MWPM for the planar codes. Positive values indicate lower logical-failure rates for the best learned decoder, while negative values indicate that the conventional decoder remains better. The heatmap shows that learned decoding does not provide a universal advantage; instead, its value depends on the code family and noise model. This supports the intended use of the framework: selective learned assistance with trusted fallback, rather than universal replacement of established decoders.

\subsection{Discussion}
\label{subsec:discussion}

The results show that pooled meta-decoding is feasible, but its reliability depends on how the decoder is evaluated. At the teacher-label level, the pooled representation is highly learnable: the Meta-MLP nearly solves interpolation and remains robust under unseen-$p$ and unseen-noise transfer. Transfer becomes harder when the code structure or code size changes, with Planar5$\times$5 consistently emerging as the main bottleneck. This suggests that shifts in error rate or noise family are easier to absorb than changes in the underlying syndrome-recovery map.

The VQC meta-decoders do not surpass the Meta-MLP in raw accuracy, so they should not be interpreted as accuracy winners. Their role is instead to test compact quantum meta-decoders under different selection criteria. The hardware-aware score changes the selected circuit architecture while keeping final accuracy close to the accuracy-selected VQC in most regimes. This supports hardware-aware VQC selection as an architecture-selection tool within the simulated search space. This interpretation is consistent with QAS as a model-selection problem rather than an automatic performance guarantee \cite{Du2022QCAS,Martyniuk2024QASSurvey}.

The key reliability finding is that teacher-label accuracy and logical reliability are not equivalent. In the hardest Planar5$\times$5 setting, raw learned decoding can remain far from teacher-level logical behavior even when classification accuracy appears acceptable. Confidence-gated fallback changes this behavior by using the learned decoder only on high-confidence cases and routing uncertain cases to the teacher decoder. The resulting system is therefore a selective learned-assisted decoder, not a full replacement for MWPM or other trusted decoders. This supports a QEC-decoder perspective in which logical-level behavior, rather than supervised prediction accuracy alone, guides decoder assessment \cite{deMartiOlius2024SurfaceCodeReview,Fischer2023HardnessRF}.

These results define the scope of the proposed framework: hardware-aware VQC selection within a compact simulated search space, and reliability-aware learned decoding through confidence-gated fallback.

\section{Conclusion}
\label{sec:conclusion}

This work presented a selective meta-decoding framework for quantum error correction that combines pooled multi-code learning, hardware-aware VQC architecture search, and confidence-gated hybrid recovery. The framework studies transfer across code, noise, and error-rate regimes instead of training a separate learned decoder for each setting.
The results show that pooled teacher-label decoding can transfer across several QEC settings. The Meta-MLP provides the strongest raw accuracy, while the VQC meta-decoders learn transferable structure above the majority baseline and support hardware-aware circuit selection within the simulated search space.

At the logical level, the main finding is that teacher-label accuracy alone is not sufficient for decoder assessment. In the hardest Planar5$\times$5 regime, confidence-gated fallback reduces the gap between raw learned decoding and teacher-level logical behavior by using learned recovery only on high-confidence cases. The resulting method is therefore a selective learned-assisted decoder rather than an unconditional replacement for established decoders.
Future work can extend this framework through calibrated confidence thresholds, broader circuit-search spaces, and implementation-level cost measurements.


\section*{Acknowledgments}
This work was supported by the HRD Group of the Council of Scientific \& Industrial Research (CSIR) provided by the CSIR Research Fellowships, and in part by the NYUAD Center
for Quantum and Topological Systems (CQTS), funded by
Tamkeen under the NYUAD Research Institute grant CG008. 

The authors also acknowledge the National Supercomputing Mission (NSM) for providing computing resources of ``PARAM Shivay'' at the Indian Institute of Technology (BHU), Varanasi, which is implemented by C-DAC and supported by the Ministry of Electronics and Information Technology (MeitY) and Department of Science and Technology (DST), Government of India.

\bibliographystyle{unsrtnat}
\bibliography{bibliography}

\clearpage
\appendix
\onecolumn

\section{\texorpdfstring{Planar5$\times$5}{Planar5x5} Logical-Failure Curves}
\label{app:planar55_logical_curves}

\begin{figure}[!ht]
    \centering
    \includegraphics[width=0.94\textwidth]
    {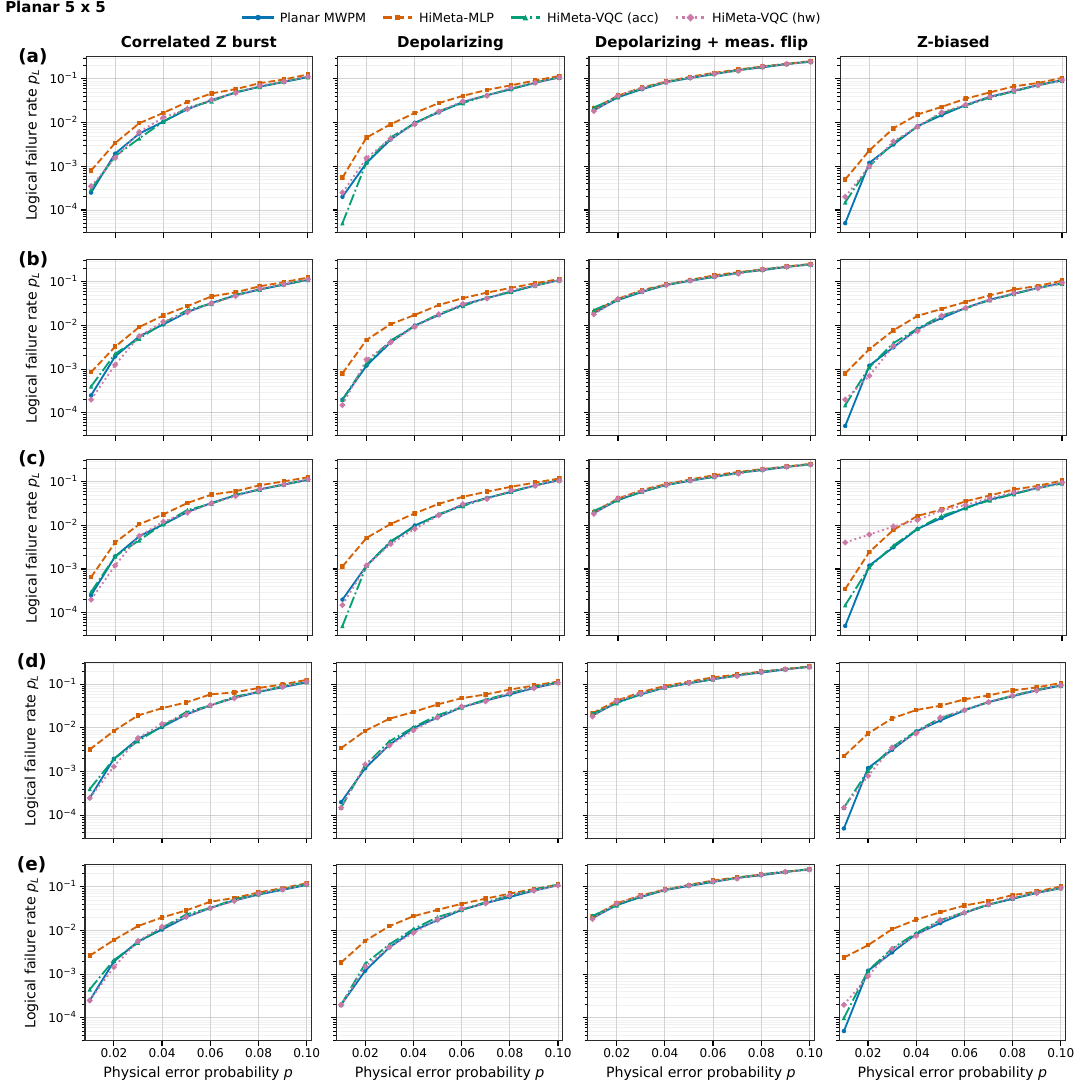}
    \caption{Additional Planar5$\times$5 logical-failure curves under the
    confidence-gated hybrid recovery protocol. Panels (a)--(e) correspond
    to interpolation, unseen-\(p\) transfer, unseen-noise transfer,
    few-shot unseen-code adaptation, and few-shot held-out-size adaptation,
    respectively.}
    \label{fig:planar55-interp-logic}
\end{figure}

The Planar5$\times$5 curves presented in
\autoref{fig:planar55-interp-logic} provide a focused analysis of the
hardest code setting and show how confidence-gated fallback changes
logical-failure behavior across the five evaluation regimes.

\clearpage
\twocolumn


\end{document}